\documentclass[]{interact}

\usepackage{epstopdf}
\usepackage[caption=false]{subfig}

\usepackage[numbers,sort&compress]{natbib}
\bibpunct[, ]{[}{]}{,}{n}{,}{,}
\renewcommand\bibfont{\fontsize{10}{12}\selectfont}
\makeatletter
\def\NAT@def@citea{\def\@citea{\NAT@separator}}
\makeatother

\theoremstyle{plain}

\theoremstyle{definition}

\theoremstyle{remark}

\usepackage{moreverb,url}

\usepackage[colorlinks,bookmarksopen,bookmarksnumbered,citecolor=red,urlcolor=red]{hyperref}

\usepackage{amsmath,amssymb,amsfonts,mathtools}
\usepackage{algorithmic}
\usepackage{graphicx}
\usepackage{textcomp}
\usepackage{xcolor}
\usepackage{subfig}
\usepackage[capitalise,nameinlink,noabbrev]{cleveref} 
\creflabelformat{equation}{#2#1#3}
\usepackage{tikz}
\usepackage{array}
\usetikzlibrary{arrows, decorations.markings}
\usetikzlibrary{decorations.pathreplacing}
\usetikzlibrary{decorations.text}
\usetikzlibrary{decorations.pathmorphing}
\usetikzlibrary{arrows}
\usetikzlibrary{patterns}
\usetikzlibrary{arrows.meta}
\newenvironment{conditions}
{\par\vspace{\abovedisplayskip}\noindent\begin{tabular}{>{$}l<{$} @{${}={}$} l}}
	{\end{tabular}\par\vspace{2\belowdisplayskip}}

\immediate\write18{texcount -inc -incbib 
	-sum borra.tex > /tmp/wordcount.tex}

\immediate\write18{texcount -char -freq
	borra.tex > /tmp/charcount.tex}

\begin{document}


\title{Development and Simulation-based Testing of a 5G-Connected Intersection AEB System}

\author{
\name{Michael Khayyat\thanks{CONTACT Michael Khayyat. Email: michael.khayyat@polimi.it}, Stefano Arrigoni\thanks{CONTACT Stefano Arrigoni. Email: stefano.arrigoni@polimi.it}, Federico Cheli\thanks{CONTACT Federico Cheli. Email: federico.cheli@polimi.it}}
\affil{Politecnico di Milano
	Department of Mechanical Engineering
	Via La Masa 1,
	20156 Milano
	MI, Italy}
}

\maketitle

\begin{abstract}
In Europe, 20\% of road crashes occur at intersections. In recent years, evolving communication technologies are making V2V and V2I faster and more reliable; with such advancements, these crashes, as well as their economic cost, can be partially reduced. In this work, we concentrate on straight path intersection collisions. Connectivity-based algorithms relying on 5G technology and smart sensors are presented and compared to a commercial radar AEB logic in order to evaluate performances and effectiveness in collision avoidance or mitigation. The aforementioned novel safety systems are tested in a blind intersection and low adherence scenario. The first algorithm proposed is obtained by incorporating connectivity information to the original control scheme, while the second algorithm proposed is a novel control logic fully capable of utilizing also adherence estimation provided by smart sensors. Test results show an improvement in terms of safety for both the architecture and high prospects for future developments.
\end{abstract}


\section{Introduction}
According to the World Health Organization (WHO), road traffic injuries are the 8th leading cause of death in the world, and the leading cause of death for children and young adults aged 5-29 years~\cite{world2018global}. The main cause of death from road traffic accidents is the inappropriate behavior of drivers, such as: speeding, driving under the influence of alcohol or drugs, distracted driving, as well as non-use of seat-belts or safety equipment. Also, some accidents are due to bad road infrastructure design (such as blind zones) that does not provide the driver with enough information to make the correct safe decision at the right time. Although human safety is the primary concern, it is worth mentioning that fatal road crashes affect the economy as well, costing up to $\$280$ billion in the United States \cite{lombardi2017age} with a similar number in Europe \cite{sander2018market}.

With 20\% of all road traffic fatalities in the European Union and United States occurring at intersections, their safety is becoming a concern \cite{sander2018market}. What is intriguing, and yet disappointing, is the fact that nearly all accidents occur at signal-controlled (52\%) or stop-controlled (46\%) intersections \cite{scanlon2017injury}. It is agreed that Connected Advanced Driver Assistance Systems (C-ADAS) are needed to help drivers navigate safely through intersections. In fact, a study done by Scanlon \textit{et al.} predicted that Intersection ADAS that delivers a warning only (no Autonomous Emergency Braking (AEB)) could prevent up to 23\% of intersection crashes, while one that brakes as well as warns could prevent up to 79\% of crashes \cite{scanlon2017injury}. It could be concluded that improving intersection safety does not only save lives, but also reflects well on the economy.

\section{Literature Review}
Vehicle-to-vehicle (V2V) and Vehicle-to-infrastructure (V2I) technologies are bringing road safety to a new level. These technologies are expanding the scope of  ``preventable collisions" to include blind-zone intersections. The upcoming 5G technology is rendering current AEB systems outdated; this is because state-of-the-art AEB is unable to utilize the full potential of the quick and rich information that 5G is able to transmit as well as that provided by smart sensors (i.e Pirelli Cyber Tyre).
 
Forward Collision AEB is explored very well in literature. This is not the case for Intersection AEB (I-AEB), as it is a relatively new concept. Euro NCAP prioritizes I-AEB development and plans on incorporating it in most vehicles by 2020 \cite{euro2015euro}. It is noteworthy that Sander \textit{et al.} explore various clustering  methods to define a small number of representative intersection test scenarios, in order to validate any  AEB logic developed as well as reduce the number of intersection scenarios tested due to their diverse nature \cite{sander2018potential}. Although no set of clusters that could group accidents into homogeneous groups was found, this paper was very insightful; through the results, it becomes clear that using simulation software is an inevitable part of testing and development, as physical testing will be very costly, in terms of money and time.
\subsection{Autonomous Emergency Braking Algorithms}
The main characteristic of current AEB algorithms is having a {predefined static braking logic} which is triggered or activated by a certain collision risk index: a quantitative measure about driving safety. While this reduces the complexity of the system, it is not suitable for all scenarios.

\subsubsection{Collision Risk Index -- Time-to-collision}
At first, Time-to-collision (TTC) is considered as it is the most widely used indicator. Usually, TTC is calculated using information given by the radar system using a constant velocity or a constant acceleration vehicle model, and after that, if the TTC is less than a certain threshold a warning is issued and if the driver doesn't respond, autonomous braking takes place based on a certain braking logic. Jeon \textit{et al.} develop a method to take into consideration the road condition in the AEB logic. As mentioned earlier, the AEB was activated if the calculated TTC is less than a certain pre-defined threshold (usually 2 seconds). However, to consider various road conditions, the threshold was modified such that it increases as the ground friction decreases (the ground friction dictates the maximum deceleration that the vehicle could achieve), as seen in Equation~\ref{eq:33x}:
\begin{equation}
	\label{eq:33x}
	\text{TTC}_{\text{threshold}} = \frac{V_{\text{ego}}}{2\mu g}
\end{equation}
where $V_{\text{ego}}$ is longitudinal ego-vehicle velocity; ${\mu}$ is ground friction (imposed as a fixed parameter) and $g$ is acceleration due to gravity. 
The latter AEB logic was experimentally tested, and showed promising results. However, the main disadvantage of such technique is that full braking is employed once the $\text{TTC}$ threshold is crossed, which is not very comfortable for the driver and not always safe \cite{jeon2016improved}.

\subsubsection{Collision Risk Index -- Time-to-react}
Time-to-react or TtR is defined as the remaining time until the very last possible driving maneuver that could avoid an imminent collision \cite{tamke2011flexible}. For example, if the possible driving maneuvers considered are: Time-to-brake (TTB) and Time-to-steer (TTS), then $\text{TtR} = \max (\text{TTB,TTS})$.

\subsubsection{Collision Risk Index -- Braking Distance}
An interesting work incorporating this index was developed by Lee \textit{et al.}. The braking distance calculated takes into consideration the effects of the actuator dynamics of the braking system leading to a better prediction \cite{lee2014development}, as in Equation~\ref{eq:6}

\begin{equation}
	\label{eq:6}
	d_{\text{brake}}= \frac{v_x^2}{2a_x} + \frac{v_x}{2J_{\text{act}}}a_x - \frac{1}{24J_{\text{act}}^2}a_x^3
\end{equation}
Where:
\begin{conditions}
	v_x & current vehicle speed \\
	a_x & desired deceleration \\
	J_{\text{act}} & actuator response time
\end{conditions}

It is also interesting to note that Seiler \textit{et al.} highlight other types of AEB logics which use braking distance as the safety indicator \cite{seiler1998development}. Mazda's algorithm \cite{doi1994development} (developed for Forward AEB) defines the braking distance as obtained in Equation~\ref{eq:7}:
\begin{equation} 
	\label{eq:7}
	d_{br} = \frac{1}{2}\bigg(\frac{v^2}{\alpha_1} - \frac{(v-v_{\text{rel}})^2}{\alpha_2}\bigg) + v\tau_1 + v_\text{rel} \tau_2 + d_0
\end{equation}
Where:
\begin{conditions}
	v & ego vehicle velocity\\
	v_\text{rel} & $v-v_\text{preceding}$\\
	\alpha_1 & maximum deceleration of the ego vehicle\\
	\alpha_2 & maximum deceleration of the preceding vehicle\\
	\tau_1,\ \tau_2 & delay time\\
	d_0 & headway offset
\end{conditions}
The Mazda system issues a warning when the range is less than braking distance plus a parameter ($d_{br} +\epsilon$) where $\epsilon$ is the system parameter. The brakes are applied when the range is less than braking distance ($d_{br}$). This systems attempts to avoid all collisions, even the extreme cases \cite{seiler1998development}; hence, warnings are issued frequently. It would be interesting to note that in such systems, drivers were seen to become desensitized to warnings as such conservative braking distance definition intervened with normal driving scenarios. Honda's algorithm  is also similar to Mazda's algorithm and is explored in~\cite{fujita1995radar}.

It is also interesting to mention the work of Malinverno \textit{et al.} where they assess the performance of a V2I collision avoidance system \cite{malinverno2018performance}. The proposed system relies on two parameters for judgment: Time-to-collision and braking distance. It is argued that these two parameters have a huge impact on the performance of the system. For example, relaxing them will lead to a system that triggers many alerts. On the other hand, being strict will lead to a system where collisions are not detected or may be detected late for the AEB to be effective. It is also noteworthy that their work is applicable for both human driven vehicles as well as autonomous ones. Testing was carried out in a virtual simulation environment to test the reliability of the algorithm; these tests showed promising results.

It is noteworthy to note that in literature, most work done on Connected Intersection AEB systems is based on a 4G network. The characteristics of such network reduces the scope of the applicability of the work as the mean communication delay of such systems is noted in literature to be in the order of hundrends of milliseconds ~\cite{amjad2018latency} and \cite{fettweis2014tactile}, which is much higher than milliseconds required~\cite{instance12x90}.

\subsubsection{Braking Logics}

In literature, there are many approaches to the implementation of the braking logic. In some cases, the braking logic depends on both the velocity and TTC, as an example \cite{kim2015aeb} is illustrated in Figure~\ref{fig:alt}. The thresholds reported are calculated as follows:
\begin{conditions}
	t_{LPB} & $\displaystyle -\frac{v_{rel}}{2\mu g}$ (Last Point to Brake) \\[2ex]
	t_{LPS} & $\displaystyle  \sqrt{\frac{2s_y}{\mu g}}$ (Last Point to Steer) 
\end{conditions}
\begin{figure}
	\centering
	\scalebox{0.6}{
		\begin{tikzpicture}
			\draw[fill = blue!30]  (-1,2) rectangle (-4,3.5) node [pos = 0.5, text width =3 cm, align = center] {\textbf{Velocity Region}};
			\draw[fill = blue!30]  (1.5,2) node (v1) {} rectangle (-1,3.5)node [pos = 0.5, text width =3 cm, align = center] {\textbf{Collision \\  Level}};
			\draw[fill = blue!30]   (8,2) rectangle (1.5,3.5)node [pos = 0.5, text width =3 cm, align = center] {\textbf{Brake Profile}};
			\draw[fill = yellow!30]   (-4,2) rectangle (-1,0.5)node [pos = 0.5, text width =2 cm, align = center] {LOW};
			\draw[fill = yellow!30]   (1.5,0.5) rectangle (-1,2)node [pos = 0.5, text width =2 cm, align = center] {Avoidance};
			\draw[fill = yellow!30]   (v1) rectangle (5.5,0.5)node [pos = 0.5, text width =3 cm, align = center] {$\text{TTC}<t_{LPB}$};
			
			\draw[fill = red!30]  (-4,0.5) rectangle (-1,-2.5)node [pos = 0.5, text width =2 cm, align = center] {HIGH};
			\draw[fill = red!30]  (1.5,-2.5) node (v4) {} rectangle (-1,0.5)node [pos = 0.5, text width =2 cm, align = center] {Mitigation};
			\draw[fill = red!30]  (5.5,-1) rectangle (1.5,0.5)node [pos = 0.5, text width =4 cm, align = center] {$t_{LPS}<\text{TTC}<t_{LPB}$};
			\draw[fill = red!30]  (5.5,-2.5) rectangle (1.5,-1)node [pos = 0.5, text width =3 cm, align = center] {$\text{TTC}<t_{LPS}$};
			\draw[fill = yellow!30] (8,0.5) rectangle (5.5,2)node [pos = 0.5, text width =3 cm, align = center] {Full Brake};
			\draw[fill = red!30]  (8,-1) rectangle (5.5,0.5)node [pos = 0.5, text width =3 cm, align = center] {Pre-Brake};
			\draw[fill = red!30]  (8,-2.5) rectangle (5.5,-1)node [pos = 0.5, text width =3 cm, align = center] {Full Brake};
	\end{tikzpicture}}
	\caption[Alternative Braking Logic]{Alternative Braking Logic \cite{kim2015aeb}}
	\label{fig:alt}
\end{figure}

\noindent where $s_y$ represents relative distance between the vehicles and two velocity regions are considered. 
\noindent On the other hand, the AEB braking logic developed in \cite{cho2014usability} is TTC dependent:\\[1ex]
\begin{tabular}{@{\tiny$\bullet$\hspace{2pt}}ll}
	if $ 1.6 \text{s} < \text{TTC} \leq 2.0 \text{s}$      & $\rightarrow \text{dec}_{\text{req}} = - 3 \text{m/s$^2$} $\\
	if $ 0.7 \text{s} < \text{TTC} \leq 1.6 \text{s}$ & $ \rightarrow \text{dec}_{\text{req}}= - 6 \text{m/s$^2$} $\\
	if $ \text{TTC} \leq 0.7 \text{s}$   & $\rightarrow$ $\text{dec}_{\text{req}} = - 10 \text{m/s$^2$} $
\end{tabular}\\[1ex]
where $\text{dec}_{\text{req}}$ represents deceleration requested.
In their work, Kapse \textit{et al.} present a detailed Simulink implementation of an AEB algorithm. The used AEB logic implements cascaded braking, which consists of two partial braking stages and one full braking stage. It is interesting to note that the AEB logic (transition between stages) does not depend on pre-defined conditions, but rather on dynamic online-calculated thresholds \cite{kapse2019implementing}. These thresholds are velocity dependent and defined based on braking times, as follows:
\begin{equation*}
	t_{\text{threshold,a}} = \frac{V_{\text{ego}}}{\text{dec}_\text{a}}, \ \
	t_{\text{threshold,b}} = \frac{V_{\text{ego}}}{\text{dec}_\text{b}}, \ \ 
	t_{\text{threshold,max}} = \frac{V_{\text{ego}}}{\text{dec}_\text{max}}
\end{equation*}      
where $\text{dec}_{a} < \text{dec}_{b} < \text{dec}_\text{max}$ and are three predefined deceleration values. These parameters are combined to obtain the three stages braking maneuver:\\[1ex]
\begin{tabular}{@{\tiny$\bullet$\hspace{2pt}}ll}
	$\text{TTC} <  t_{\text{threshold,a}}$      & $\rightarrow\text{dec}_{\text{req}} = \text{dec}_\text{a}$\\
	$t_{\text{threshold,a}} \leq\text{TTC} <  t_{\text{threshold,b}}$ & $\rightarrow\text{dec}_{\text{req}} = \text{dec}_\text{b}$\\
	$t_{\text{threshold,b}} \leq\text{TTC} <  t_{\text{threshold,max}}$   & $\rightarrow\text{dec}_{\text{req}} = \text{dec}_\text{max}$
\end{tabular}

\section{Description of the Setup}
\subsection{Current System Architecture}
\label{sec:currenxt}
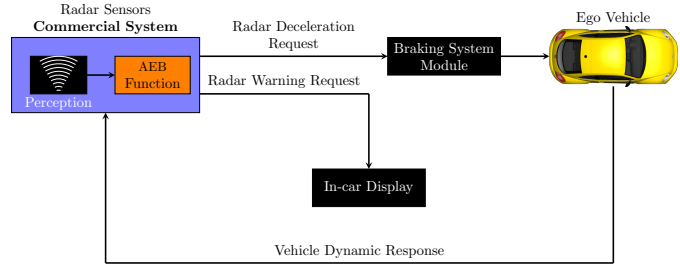
\begin{figure}[h]
	\centering
	\scalebox{0.5}{
		\begin{tikzpicture}
			\node[rotate=180] at (0,0.5) {\includegraphics[scale=0.3]{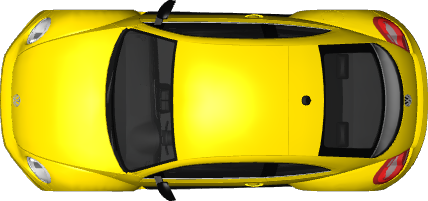}};
			\node at (0,1.5) {Ego Vehicle};
			\draw[fill=blue!50]  (-16,1) rectangle (-11,-1);
			\draw[fill=black]  (-15.5,0.5) rectangle (-14,-0.5);
			\draw[fill=orange]   (-13.25,0.5) rectangle (-11.25,-0.5)node[pos=.5,align=center]{\small AEB \\ Function};
			\draw [decorate,decoration={expanding waves, angle=30,segment length=1mm},white](-14.75,-0.5) -- (-14.75,0.5);
			\node[white] at (-14.75,-0.75) {Perception};
			\node[text width = 5cm,align = center] at (-13.5,1.5) {Radar Sensors \\ \textbf{Commercial System}};
			\draw[fill = black]  (-6,1) rectangle (-3,0)node[pos=.5,align=center,white]{Braking System\\ Module};
			\draw[-stealth,very thick] (-11,0.5) -- (-6,0.5)node[pos=.5,above,text width = 5.5cm,align = center]{Radar Deceleration\\Request };
			\draw[-stealth,very thick]  (-3,0.5) -- (-1.7,0.5);
			\draw[fill = black]    (-8,-2.5) rectangle (-5,-3.5)node[pos=.5,align=center,white]{In-car Display};
			\draw[very thick] (-11,-0.5) -- (-6.5,-0.5)node[pos=.5,above]{Radar Warning Request}; 
			\draw[-stealth,very thick](-6.5,-0.5) -- (-6.5,-2.5);
			\draw[very thick] (-0,-0.3)--(0,-5) ;
			\draw[very thick] (0,-5) -- (-13.5,-5)node[pos=.5,above]{Vehicle Dynamic Response };
			\draw[-stealth,very thick] (-13.5,-5)-- (-13.5,-1);
			\draw[-stealth,very thick] (-14,0) -- (-13.25,0);
	\end{tikzpicture}}
	\caption[Commercial System Architecture]{Commercial System Architecture}
	\label{fig:currentArch}
\end{figure}

The architecture of the current commercial system considered in this work is shown in Figure~\ref{fig:currentArch}. 

The ``Radar Sensors" block is responsible for perception, computation of TTC, and finally issuing the deceleration request to the braking system module, which in turn regulates the brakes pressure to reach the desired deceleration.

At first, information coming from the Radar Sensors about surrounding vehicles is evaluated to determine if the TTC should be calculated or not. The \textit{activation criteria} for the AEB feature is the following:
\begin{itemize}
	\item Target vehicles with orthogonal path: $45$\textdegree $<|\arctan\frac{v_y}{v_x}| <$ $135$\textdegree
	\item Target vehicle with $|v_y|> 2 \text{ m/s}$
	\item Ego vehicle with $5 \text{ km/h}\leq v_{ego} \leq  60 \text{ km/h}$
\end{itemize}
where $v_y$,$v_x$ are respectively the target vehicle longitudinal and lateral velocities.
It is important to note that the AEB is intended to operate in \textit{urban areas}. The choice of limiting the orthogonal path of the target vehicle is explained by the fact that the operation of the system is limited to intersection application. For example, the AEB system should not be activated when doing overtakes on a road.

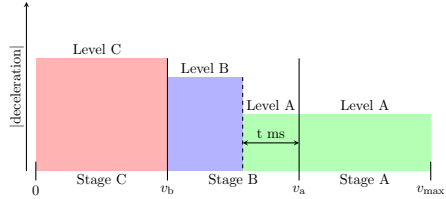
\begin{figure}[h]
	\centering
	\scalebox{0.5}{
		\begin{tikzpicture}
			\node at (-5,-3) {$0$};
			\node at (5.5,-3)  {$v_\text{max}$};
			\node at (2,-3)  {$v_\text{a}$};
			\node at (-1.5,-3) {$v_\text{b}$};
			\draw[fill = green!30,draw = none] (5.5,-1) rectangle (2,-2.5) ;
			\draw[fill = green!30,draw = none]   (2,-1) rectangle (0.5,-2.5) node (v1) {};
			\draw[fill = blue!30,draw = none]  (0.5,-2.5) rectangle (-1.5,0);
			\draw[fill =red!30,draw = none] (-1.5,-2.5) rectangle (-5,0.5);

			\draw[thick] (-5,-2.25) -- (-5,-2.75);
			\draw[thick] (5.5,-2.25) -- (5.5,-2.75);
			\draw[thick] (-1.5,0.5) -- (-1.5,-2.75);
			\draw[thick](2,0.5) -- (2,-2.75);
			
			\draw[thick,-stealth] (-5.25,-2.5) -- (-5.25,2);
			\node[rotate = 90] at (-5.5,-0.25) {$|\text{deceleration}|$};
			\node at (-3.375,0.75) {Level C};
			\node at (-0.5,0.25) {Level B};
			\node at (1.25,-0.75) {Level A};
			\node at (3.75,-0.75) {Level A};
			\node at (0.25,-2.75) {Stage B};
			\node at  (3.75,-2.75) {Stage A};
			\node at (-3.25,-2.75) {Stage C};
			\draw[thick,dashed] (0.5,0) -- (0.5,-2.5);
			\draw[thick,stealth-stealth] (0.5,-1.75) -- (2,-1.75) node [pos = 0.5,above]{t ms};
	\end{tikzpicture}}
	\caption[Scheme of Commercial AEB Logic]{Scheme of Commercial AEB Logic}
	\label{fig:scheme}
\end{figure}

The sensing architecture is composed only of 3 radar sensors, as reported in Figure \ref{fig:setupsens} as well as a measurement of the ego vehicle velocity as provided by CAN. It is noteworthy that the side radars (shown in red) are short range ones (SRR) and the center one (shown in blue) is a long range radar (LRR).
The Radar Deceleration Request is issued once the calculation of TTC calculated by the ``Perception" Block is $<1.5$ s. In this work, the proposed commercial AEB logic provided by the ``AEB Function" block in Figure \ref{fig:currentArch} is velocity-dependant. 
\begin{itemize}
	\item if $v \leq  v_{\text{b}} \rightarrow$  $\text{dec}_{\text{requested}} = \text{Level C}  \text{ m/s$^2$}$
	\item if $   v_{\text{b}}  <v \leq  v_{\text{a}} \rightarrow$ $\text{dec}_{\text{requested}} = \text{Level A}  \text{ m/s$^2$}$ for $300$ ms, followed by $\text{dec}_{\text{requested}} = \text{Level B}  \text{ m/s$^2$}$
	\item if $  v_{\text{a}}  <v \leq  v_{\text{max}}\rightarrow$ $\text{dec}_{\text{requested}} = \text{Level A } \text{m/s$^2$}$
\end{itemize}
It is important to note that value of the deceleration levels could not be provided at the request of our research partners. For this purpose, a scheme is provided in Figure~\ref{fig:scheme} as an additional support to fully understand the logic.
\subsection{Connected System Architecture}
\begin{figure}[h]
	\centering
	\scalebox{0.475}{
		\begin{tikzpicture}
			\node[rotate=180] at (-0.6375,0.5) {\includegraphics[scale=0.3]{img/vwtop.png}};
			\node at (-0.6375,1.5) {Ego Vehicle};
			\draw[fill=blue!50]  (-17.75,1) rectangle (-13,-1);
			\draw[fill=black]  (-17.5,0.5) rectangle (-16,-0.5);
			\draw[fill=orange]   (-15.25,0.5) rectangle (-13.25,-0.5)node[pos=.5,align=center]{AEB \\ Function};
			\draw [decorate,decoration={expanding waves, angle=30,segment length=1mm},white](-16.75,-0.5) -- (-16.75,0.5);
			\node[white] at (-16.75,-0.75) {Perception};
			\node[text width = 5cm,align = center] at (-15.375,1.5) {Radar Sensors \\ \textbf{Commercial System}};
			\draw[fill = black]  (-5.75,1) rectangle (-2.75,0)node[pos=.5,align=center,white]{Braking System\\ Module};
			\draw[very thick] (-13,0.5) -- (-8,0.5)node [pos=.5,above] (v6) {\ \ Radar Deceleration Request};
			\draw[-stealth,very thick]  (-2.75,0.5) -- (-2.25,0.5);
			\draw[fill = black]    (-7.5,-5.5) rectangle (-4.5,-6.5)node[pos=.5,align=center,white]{In-car Display};
			\draw[very thick] (-13,-0.5) -- (-8.5,-0.5)node [pos=.5,above] (v5) {\ \ Radar Warning Request}; 
			\draw[very thick] (-0.6375,-0.3)--(-0.6375,-8) ;
			\draw[very thick] (-0.6375,-8) -- (-15.375,-8)node[pos=.5,above]{Vehicle Dynamic Response };
			\draw[-stealth,very thick] (-15.375,-8)-- (-15.375,-1);
			\draw[-stealth,very thick] (-16,0) -- (-15.25,0);
			\draw[fill=blue!50] (-14.5,-6) rectangle (-12.5,-4);
			\node[text width = 3cm,align = center]  at (-13.5,-3.5)  {5G \\ Virtual Sensor};
			\node[white] at (-13.5,-5.5) {Perception};
			\draw  (-13.5,-4.375) node (v3) {} ellipse (0.125 and 0.0625);
			\draw (-13.625,-4.375) -- (-13.75,-5.25) node (v1) {} ;
			\draw (-13.25,-5.25) node (v2) {} -- (-13.375,-4.375);
			\draw  plot[smooth, tension=.7] coordinates {(v1) (-13.5,-5.3125) (v2)};
			\draw   (-13.5,-4.375) -- (-13.5,-4.25) node (v4) {};
			\draw  (v4) ellipse (0.03125 and 0.03125);
			\draw [decorate,decoration={expanding waves, angle=80,segment length=0.3mm}] (-13.5,-4.25) -- (-13.5,-4.0625);
			
			\draw[fill=green!50!black]  (-11,-1.5) rectangle (-6.75,-3.5);
			\draw[fill=green!80!black]  (-10.75,-2) rectangle (-9,-3)node[pos=.5,align=center,white]{5G AEB \\ Function};
			\draw[fill=green!80!black]  (-8.5,-2) rectangle (-7,-3)node[pos=.5,align=center,white]{AEB\\Merge };
			\draw[-stealth,very thick]  (-9,-2.5) -- (-8.5,-2.5);
			
			\draw[-stealth,very thick]  (-8.5,-0.5) -- (-8,-0.5) -- (-8,-1.5);
			\draw[-stealth,very thick] (-8,0.5) -- (-7.5,0.5) -- (-7.5,-1.5);
			\draw[very thick] (-6.75,-2.25) -- (-6.25,-2.25) node (v7) {};

			\draw[very thick] (-6.25,-2.25)-- (-6.25,0.5) node (v8) {};
			
			\draw[-stealth,very thick] (-6.25,0.5)-- (-5.75,0.5);
			\node at (-8.75,-3.75) {\textbf{MABX ECU}};
			\draw[very thick](-6.75,-2.75) -- (-6.25,-2.75) node (v9) {};
			
			\draw[-stealth,very thick](-6.25,-2.75)-- (-6.25,-5.5);
			\draw[very thick] (-12.5,-5) -- (-11.5,-5) node (v10) {};
			\draw[very thick] (-11.5,-5) -- (-11.5,-2.5) node (v11) {};
			\draw[-stealth,very thick](-11.5,-2.5) -- (-11,-2.5);
			
			\draw[-stealth,very thick](-13.5,-8) -- (-13.5,-6);
			\draw[fill = black!30]  (-14.5,-1.5) rectangle (-12.5,-2.5);
			\node at (-13.5,-1.75) {\includegraphics[scale=0.3]{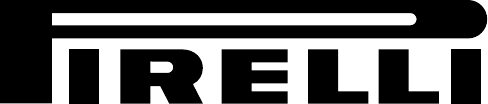}};
			\node at (-13.5,-2.25) {Cyber Tyre};
			
			\draw[-stealth,very thick](-12.5,-2) -- (-11,-2);
	\end{tikzpicture}}
	\caption[Connected System Architecture]{Connected System Architecture}
	\label{fig:modifiedArch}
\end{figure}

The architecture considered for the development of the novel connected AEB algorithms is shown in Figure~\ref{fig:modifiedArch}.

It is important to note that the operation of the radar safety system remains the same (i.e activation criteria and braking logic) as that described in the previous section. It could be seen in Figure~\ref{fig:setupsens} that the additional sensors mounted on the vehicle of the connected system are the following: 
\begin{figure}[h]
	\usetikzlibrary{decorations.pathreplacing}
	\centering
	\scalebox{0.6}{
		\begin{tikzpicture}
			
			\node[rotate=-90] (v1) at (0,0) {\includegraphics[scale=0.5]{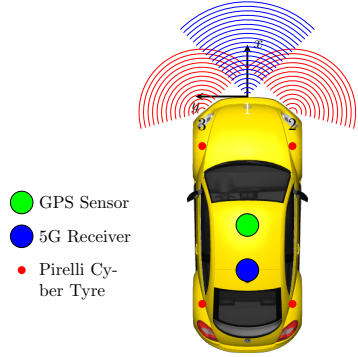}};
			\draw[very thick, -stealth] (0,2.85) node[below,white] (v3) {1} -- (0,4);
			\draw[very thick, -stealth] (0.02,2.85) node (v2) {} -- (-1.15,2.85);
			\node[right] at (0,4) {$x$};
			\node[below] at (-1.15,2.85) {$y$};
			
			\draw [rotate around={-30:(1,2.5)}, decorate,decoration={expanding waves, angle=75,segment length=1mm},red](1,2.5)node[below,black] (v3) {2}  -- (1,4);
			\draw [rotate around={30:(-1,2.5)},decorate,decoration={expanding waves, angle=75,segment length=1mm},red](-1,2.5)node[below,black] (v3) {3}   -- (-1,4);
			\draw[decorate,decoration={expanding waves, angle=45,segment length=1mm},blue] (0,2.85)  -- (0,5);
			
			\draw[fill=green]  (v1) ellipse (0.25 and 0.25);
			
			\draw[fill=blue]  (0,-1) ellipse (0.25 and 0.25);
			\node[circle, inner sep = 0pt, fill = red,minimum size = 5pt]  at (-1,1.75) {};
			\node[circle, inner sep = 0pt, fill = red,minimum size = 5pt] at (1,1.75) {};
			\node[circle, inner sep = 0pt, fill = red,minimum size = 5pt] at (-1,-1.75) {};
			\node[circle, inner sep = 0pt, fill = red,minimum size = 5pt] at (1,-1.75) {};
			
			\draw[fill=green]   (-5,0.5) ellipse (0.25 and 0.25);
			\draw[fill=blue]  (-5,-0.25) ellipse (0.25 and 0.25);
			\node[circle, inner sep = 0pt, fill = red,minimum size = 5pt] at (-5,-1) {};
			\node[right] at (-4.75,0.5) {GPS Sensor};
			\node[right] at (-4.75,-0.25) {5G Receiver};
			\node[right,text width = 2cm] at (-4.75,-1.25) {Pirelli Cyber Tyre};
			
	\end{tikzpicture}}
	\caption[Complete sensor configuration for connected system architecture]{Complete sensor configuration for connected system architecture}
	\label{fig:setupsens}
\end{figure}
\begin{itemize}
	\item GPS Sensor with RTK Correction (ZED-F9P) operates at 10 Hz for a global accurate localization (error $<0.5$ m).
	\item 5G Virtual Sensor receives information by other connected vehicles about their speed and location, and by the infrastructure about potential ground friction (a digital map containing potential ground friction estimated by the Cyber Tyre sensors of other vehicle who traveled on the road earlier).
	\item Pirelli Cyber Tyre is able to measure the potential ground friction \textit{after} braking takes place. 
	It is noteworthy that through the use of Kalman Filters (for slip estimation), Dynamic Estimators (for force estimation) and Pre-defined Reference Curves, the Friction-Identification Algorithm is able to provide the potential ground friction with an accuracy dependent on that of the measurements provided by the accelerometers.
\end{itemize}
\subsection{Testing Scenario}
\begin{figure}[h]
\centering
\includegraphics[scale=1]{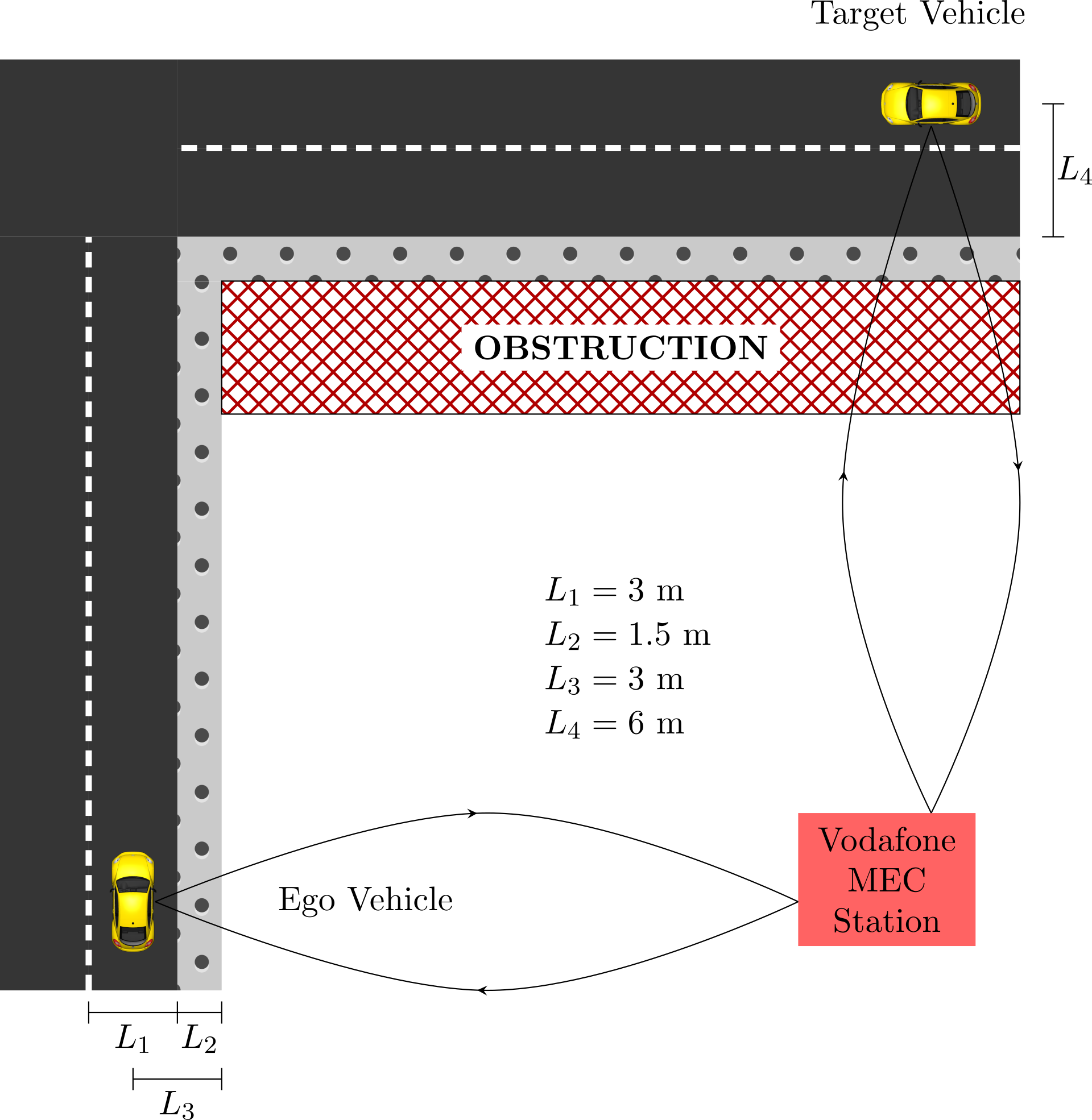}
	\caption[]{Scenario under consideration in the development of the AEB systems and their testing}
	\label{fig:most}
\end{figure}

The scenario under consideration in this work is described in Figure~\ref{fig:most}.

It is important to note that the simplified setup was chosen for two main reasons:
\begin{itemize}
	\item it is one of the most common intersection crashes schemes; 
	\item it allows the work to concentrate on the power of connectivity and its potential influence on safety systems. 
\end{itemize}
The scenario considered is a blind intersection where the initial velocity of the {Target Vehicle} is set to ensure that it crashes with the {Ego Vehicle} just before any braking takes place. It is also important to note that it is assumed that the Ego Vehicle, unlike the Target Vehicle, can be actuated.
\section{IPG CarMaker/Simulink Implementation}
IPG CarMaker is a tool for virtual testing of vehicles, which allows to recreate realistic test scenarios in a virtual environment. IPG CarMaker provides a Virtual Vehicle Environment (VVE), which consists of the virtual road, virtual driver, and virtual vehicle. Furthermore, IPG CarMaker has a user-friendly MATLAB/Simulink interface that facilitates modeling the connected system architecture, as seen in Figure \ref{fig:modifiedArch}.
\subsection{Scheme of Connected Commercial System}
\begin{figure}[h]
	\centering
	\scalebox{0.65}{
		\begin{tikzpicture}
			\draw[dashed,thick] (-0.25,1.25) node (v6) {} -- (11.25,1.25) -- (11.25,-2.25) -- (2.25,-2.25) -- (2.25,-2.75) -- (-0.25,-2.75) -- (-0.25,1.25);
			\draw[thick, loosely dashed] (-0.5,2) node (v7) {} -- (14.25,2) -- (14.25,-3.25) -- (11.75,-3.25) -- (11.75,-3.75) -- (11.25,-3.75) -- (11.25,-7.25) -- (-0.5,-7.25) -- (-0.5,2);
			\node at (7,2.25) {Connected System Model};
			\node at (7,1.5) {Current System Model};
			\draw[fill=blue!30,draw = none]  (0,0) rectangle (2,1) node [pos= 0.5,text width =3cm, align = center] {Radar \\ Sensors};
			\draw[fill=red!30,draw = none]    (0,-1.5) rectangle (2,-2.5)node [pos= 0.5,text width =3cm, align = center] {IMU};
			\draw[fill=green!30,draw = none]   (0,-4) rectangle (2,-5)node [pos= 0.5,text width =3cm, align = center] {5G \\ Module};

			\draw[fill = red]  (3,0) rectangle (5,-1.5);
			\draw[ultra thick,color = white]  (3.8632,-0.2691) arc (105.8793:275.2941:0.5);
			\draw[ultra thick,color = white]  (4.1368,-0.2691) arc (74.1207:-84.7059:0.5);
			\draw[ultra thick,color = white]   (4,-0.125) -- (4,-0.5);
			
			\draw[fill = red]  (3,-2.5) rectangle (5,-4);
			\draw[ultra thick,color = white] (3.8632,-2.7691) arc (105.8793:275.2941:0.5);
			\draw[ultra thick,color = white] (4.1368,-2.7691) arc (74.1207:-84.7059:0.5);
			\draw[ultra thick, color = white]  (4,-2.625) -- (4,-3);
			
			\node[text width =3cm, align = center] at (4,-0.75) {Activation \\ Criteria};
			\node[text width =3cm, align = center] at (4,-3.25) {Activation \\ Criteria};
			
			\draw[thick,-stealth] (2,0.5) -- (2.5,0.5) node (v2) {} -- (2.5,-0.5) -- (3,-0.5);
			\draw[thick,-stealth] (2,-2) -- (2.5,-2) node (v1) {} -- (2.5,-1) -- (3,-1);
			\draw[thick,-stealth] (2,-4.5) -- (2.5,-4.5) node (v5) {} -- (2.5,-3.5) -- (3,-3.5);
			\draw[thick,-stealth] (2.5,-2) node (v3) {} -- (2.5,-3) -- (3,-3);

			\draw[fill = blue!90!black,draw=none]  (6,0) rectangle (8,-1.5)node [pos= 0.5,text width =3cm, align = center] {TTC \\ Calculation};
			\draw[fill = blue!90!black,draw = none]  (6,-2.5) rectangle (8,-4)node [pos= 0.5,text width =3cm, align = center]{TTC/BD \\ Calculation};
			\draw[thick,-stealth] (5,-0.75) -- (5.75,-0.75);
			\draw[thick,-stealth]  (5,-3.25) -- (5.75,-3.25);
			
			\draw[thick]  (5.9375,-1.125) ellipse (0.0625 and 0.0625);
			\draw[thick]  (5.9375,-0.375) ellipse (0.0625 and 0.0625);
			\draw[thick] (5.9375,-1.0625) -- (5.6875,-0.375);
			\draw[thick]  (5.9375,-3.625) ellipse (0.0625 and 0.0625);
			\draw[thick]  (5.9375,-2.875) ellipse (0.0625 and 0.0625);
			\draw[thick] (5.9375,-3.5625) -- (5.6875,-2.875);
			\draw[{Stealth[scale=0.4]}-{Stealth[scale=0.4]}](5.5853,-0.6239) arc (128.7648:91.8462:0.5625);
			\draw[{Stealth[scale=0.4]}-{Stealth[scale=0.4]}] (5.5853,-3.1239) arc (128.7648:91.8462:0.5625);
			
			\draw[thick,-stealth] (2.5,0.5) -- (7,0.5) -- (7,0);
			\draw[thick,-stealth]  (2.5,-2) -- (7,-2) node (v4) {} -- (7,-1.5);
			\draw[thick,-stealth] (7,-2) -- (7,-2.5);
			\draw[thick,-stealth] (2.5,-4.5) -- (6.5,-4.5) -- (6.5,-4);
			
			\fill[gray!30] (9,0) rectangle (11,-1.5) node[pos = 0.5,black,text width = 2.5 cm, align = center]{Radar\\AEB};
			\fill[gray!30]  (9,-2.5) rectangle (11,-4) node[pos = 0.5,black,text width = 2 cm, align = center]{Connected\\ AEB};
			
			\draw[thick,-stealth](8,-0.75) -- (9,-0.75) ;
			\draw[thick,-stealth] (8,-3.25) -- (9,-3.25);
			\draw[draw=none, fill = green!90!black]  (12,-1.25) rectangle (14,-2.75) node[pos = 0.5] {AEB Merge};
			
			\draw[thick,-stealth] (11,-0.75) -- (11.5,-0.75) -- (11.5,-1.75) -- (12,-1.75);
			
			\draw[thick,-stealth] (11,-3.25) -- (11.5,-3.25) -- (11.5,-2.25) -- (12,-2.25);
			\draw[thick,-stealth] (13,-2.75) -- (13,-4);
			\draw[fill=red!20,draw=none]  (11.75,-4) rectangle (14.25,-5) node[pos = 0.5,text width = 2.5 cm,align = center] {Acceleration Controller};
			\draw[thick,-stealth] (13,-5) -- (13,-6);
			\node at (13,-7) {\includegraphics[scale=0.2]{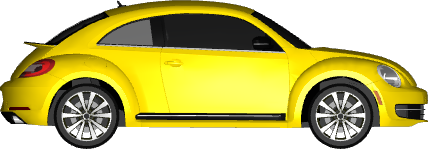}};
			\draw[fill=yellow!20,draw=none]  (0,-6) rectangle (2,-7)node[pos = 0.5,text width = 2.5 cm,align = center] {Pirelli Cyber Tyre};
			\draw[thick,-stealth]  (2,-6.5) -- (7.5,-6.5) -- (7.5,-4);
			
	\end{tikzpicture}}
	\caption[Summary of Developed System]{Summary of Developed System}
	\label{fig:developed}
\end{figure}
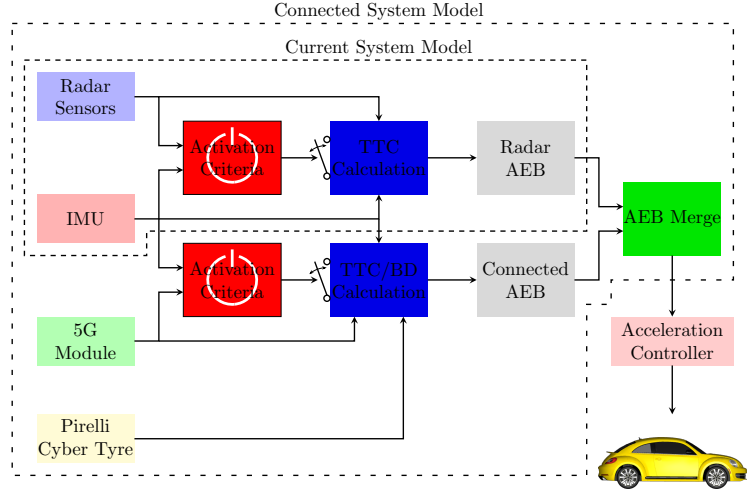

The summary of the developed Simulink model for the two control logics is presented in Figure~\ref{fig:developed}. 
\subsubsection{Current System Model}
As an initial step, the current system architecture (as modeled in Figure \ref{fig:currentArch}, and illustrated in \ref{fig:developed}) was modeled in IPG CarMaker.

For the perception block, information from the radars of the ego vehicle about surrounding visible vehicles was analyzed to determine whether to activate the TTC Calculation block or not (according to the activation criteria defined earlier).

For the purpose of this work, \cite{miller2002adaptive} and \cite{jimenez2013improved} are used to build the TTC calculation block in the Simulink model.
\begin{figure}[h]
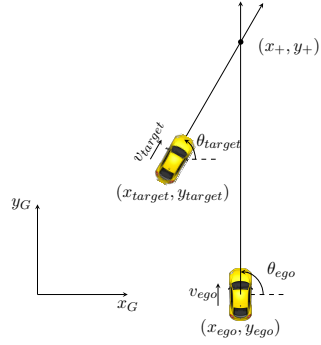

	\usetikzlibrary{patterns}
	\usetikzlibrary{calc,patterns,angles,quotes}
	
	\begin{center}
		\scalebox{0.6}{
			\begin{tikzpicture}
				\node(v1) at (3,5) {};
				\node(v2) at (4.5,2) {};
				
				\draw[-stealth](0,2) -- (0,4) node[pos=1,left] {$y_G$};
				\draw[-stealth](0,2) -- (2,2)node[pos=1,below] {$x_G$};
				
				\draw[dashed] (3,5) -- (4,5);
				\draw[dashed](4.5,2) -- (5.5,2);
				\draw[rotate around={-120:(3,5)},-stealth](3,5) -- (-1,5);			
				\node[rotate=-120] (v1) at (3,5) {\includegraphics[scale=0.1]{img/vwtop.png}};
				\node[rotate=-90] (v2) at (4.5,2) {\includegraphics[scale=0.1]{img/vwtop.png}};
				\node(k) at (3.5844,6.0072) {};
				\node(k2) at (5,5) {};
				\pic [draw, -stealth, "\hspace{0.75cm}$\theta_{target}$", angle eccentricity=1.5] {angle = k2--v1--k};
				\node(k3) at (4.5,3) {};
				\node(k4) at (4.5,2) {};
				\node(k5) at (5.5,2) {};
				\pic [draw, -stealth, "\hspace{0.75cm}$\theta_{ego}$", angle eccentricity=1.5] {angle = k5--k4--k3};
				\draw[-stealth] (4.5,2) -- (4.5,8.5);

				\node at (3,4.25) {$(x_{target},y_{target})$};
				\node at (4.5,1.25) {$(x_{ego},y_{ego})$};
				
				\fill (4.5033,7.6) ellipse (0.06 and 0.06);
				\node[right] at (4.75,7.5) {$(x_+,y_+)$};
				
				\draw[-stealth] (4,1.75) -- (4,2.25);
				\node at (3.675,2) {$v_{ego}$};
				\draw[rotate around={-30:(2.5,5)},-stealth] (2.5,5) -- (2.5,5.5);
				\node[rotate=60] at (2.5,5.375) {$v_{target}$};
		\end{tikzpicture}}
	\end{center}
	\caption[TTC in X-Y Plane]{TTC in X-Y Plane}
	\label{fig:ttcradar}
\end{figure}

\begin{figure}[h]
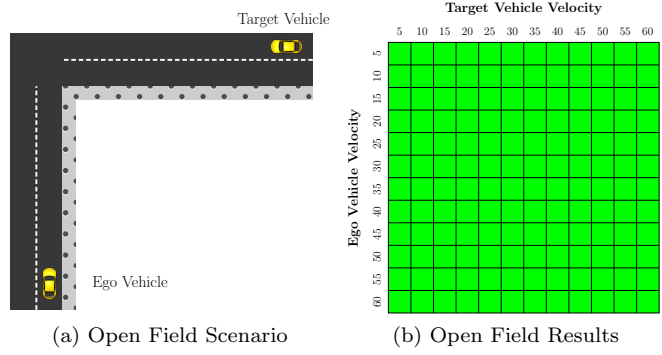

	\centering
	\subfloat[Open Field Scenario]{\centering
		\definecolor{way}{HTML}{373737}
		\scalebox{0.35}{
			\begin{tikzpicture}
				\fill[way]  (0.5,-10) rectangle (-0.5,-1.5) node (v2) {};
				\fill[way]    (-0.5,-10) node (v1) {} rectangle (-1.5,-1.5);
				\fill[way] (10,0.5) rectangle (0.5,-0.5);
				\fill[way]   (10,-0.5) node (v3) {} rectangle (0.5,-1.5);
				\draw[white,line width=2pt, dash pattern=on 5pt off 3pt,](-0.5,-10)  -- (-0.5,-1.5);
				\fill[pattern=crosshatch dots gray]  (1,-10) rectangle (0.5,-2) node (v4) {};
				\fill[pattern=crosshatch dots gray]  (10,-1.5) rectangle (v4);
				\node[rotate=-90] at (0,-9) {\includegraphics[scale=0.1]{img/vwtop.png}};
				\node[rotate=0] at (9,0) {\includegraphics[scale=0.1]{img/vwtop.png}};
				\draw[white,line width=2pt, dash pattern=on 5pt off 3pt,] (10,-0.5) -- (0.5,-0.5);
				
				\fill[way] (-1.5,0.5) rectangle (0.5,-1.5);

				
				
				
				
				\node[right] at (1.5,-9) {\LARGE Ego Vehicle};
				\node[right] at (7,1) {\LARGE Target Vehicle};
		\end{tikzpicture}}
		\label{fig:resultsof}}
	\subfloat[Open Field Results]{\scalebox{0.3}{
			\begin{tikzpicture}
				
				\draw[fill=green]  (0,0) rectangle (12,12);
				\draw (1,12) -- (1,0);
				\draw (2,12) -- (2,0);
				\draw (3,12) -- (3,0);
				\draw (4,12) -- (4,0);
				\draw (5,12) -- (5,0);
				\draw (6,12) -- (6,0);
				\draw (7,12) -- (7,0);
				\draw (8,12) -- (8,0);
				\draw (9,12) -- (9,0);
				\draw (10,12) -- (10,0);
				\draw (11,12) -- (11,0);
				
				\draw (0,11) -- (12,11);
				\draw (0,10) -- (12,10);
				\draw (0,9) -- (12,9);
				\draw (0,8) -- (12,8);
				\draw (0,7) -- (12,7);
				\draw (0,6) -- (12,6);
				\draw (0,5) -- (12,5);
				\draw (0,4) -- (12,4);
				\draw (0,3)-- (12,3);
				\draw (0,2) -- (12,2);
				\draw (0,1) -- (12,1);
				
				\node at (6,13.5) {\LARGE  \textbf{Target Vehicle Velocity}};
				\node[rotate = 90] at (-1.5,6) {\LARGE  \textbf{Ego Vehicle Velocity}};
				\node at (0.5,12.5) {\Large 5};
				\node at (1.5,12.5) {\Large 10};
				\node at (2.5,12.5) {\Large  15};
				\node at (3.5,12.5) {\Large  20};
				\node at (4.5,12.5) {\Large  25};
				\node at (5.5,12.5) {\Large  30};
				\node at (6.5,12.5) {\Large  35};
				\node at (7.5,12.5) {\Large  40};
				\node at (8.5,12.5) {\Large  45};
				\node at (9.5,12.5) {\Large  50};
				\node at (10.5,12.5) {\Large  55};
				\node at (11.5,12.5) {\Large  60};
				
				\node[rotate = 90] at (-0.5,11.5) {\Large  5};
				\node[rotate = 90] at (-0.5,10.5) {\Large  10};
				\node[rotate = 90] at (-0.5,9.5) {\Large  15};
				\node[rotate = 90] at (-0.5,8.5) {\Large  20};
				\node[rotate = 90] at (-0.5,7.5) {\Large  25};
				\node[rotate = 90] at (-0.5,6.5) {\Large  30};
				\node[rotate = 90] at (-0.5,5.5) {\Large  35};
				\node[rotate = 90] at (-0.5,4.5) {\Large  40};
				\node[rotate = 90] at (-0.5,3.5) {\Large  45};
				\node[rotate = 90] at (-0.5,2.5) {\Large  50};
				\node[rotate = 90] at (-0.5,1.5) {\Large  55};
				\node[rotate = 90] at (-0.5,0.5) {\Large  60};
		\end{tikzpicture}}
		\label{fig:openfield}}
	\caption{Description of Open Field Case}
	\label{descOF}
\end{figure}

\begin{figure}[h]
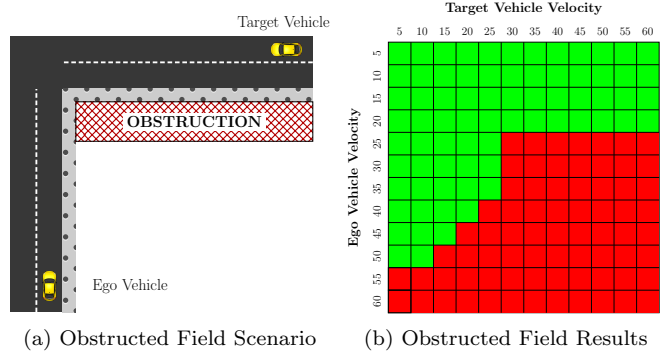

	\centering
	\subfloat[Obstructed Field Scenario]{\centering
		\definecolor{way}{HTML}{373737}
		\scalebox{0.35}{
			\begin{tikzpicture}
				\fill[way]  (0.5,-10) rectangle (-0.5,-1.5) node (v2) {};
				\fill[way]    (-0.5,-10) node (v1) {} rectangle (-1.5,-1.5);
				\fill[way] (10,0.5) rectangle (0.5,-0.5);
				\fill[way]   (10,-0.5) node (v3) {} rectangle (0.5,-1.5);
				\draw[white,line width=2pt, dash pattern=on 5pt off 3pt](-0.5,-10)  -- (-0.5,-1.5);
				\fill[pattern=crosshatch dots gray]  (1,-10) rectangle (0.5,-2) node (v4) {};
				\fill[pattern=crosshatch dots gray]  (10,-1.5) rectangle (v4);
				\node[rotate=-90] at (0,-9) {\includegraphics[scale=0.1]{img/vwtop.png}};
				\node[rotate=0] at (9,0) {\includegraphics[scale=0.1]{img/vwtop.png}};
				\draw[white,line width=2pt, dash pattern=on 5pt off 3pt,] (10,-0.5) -- (0.5,-0.5);
				\draw [pattern=crosshatch,pattern color=black!30!red] (1,-2) rectangle (10,-3.5)node[pos=.5,align=center]{};
				\node[fill=white] at (5.5,-2.75) {\LARGE  \textbf{OBSTRUCTION}};
				\fill[way] (-1.5,0.5) rectangle (0.5,-1.5);
				
				\fill[way] (-1.5,0.5) rectangle (0.5,-1.5);

				
				
				
				
				\node[right] at (1.5,-9) {\LARGE Ego Vehicle};
				\node[right] at (7,1) {\LARGE  Target Vehicle};
		\end{tikzpicture}}
		\label{fig:resultsob}}
	\subfloat[Obstructed Field Results]{\scalebox{0.3}{
			\begin{tikzpicture}
				
				\draw[fill=green]  (0,0) rectangle (12,12);
				\fill[red]  (5,8) rectangle (12,0);
				\fill[red]  (4,5) rectangle (5,4);
				\fill[red]   (3,4) rectangle (5,3);
				\fill[red]   (2,3) rectangle (5,2);
				\fill[red]    (1,2) rectangle (5,1);
				\fill[red] (1,1) rectangle (5,0);
				\fill[red,draw = black]  (0,2) rectangle (1,1);
				\fill[red,draw = black] (0,1) rectangle (1,0);
				
				\draw (1,12) -- (1,0);
				\draw (2,12) -- (2,0);
				\draw (3,12) -- (3,0);
				\draw (4,12) -- (4,0);
				\draw (5,12) -- (5,0) node (v1) {};
				\draw (6,12) -- (6,0);
				\draw (7,12) -- (7,0);
				\draw (8,12) -- (8,0);
				\draw (9,12) -- (9,0);
				\draw (10,12) -- (10,0);
				\draw (11,12) -- (11,0);
				
				\draw (0,11) -- (12,11);
				\draw (0,10) -- (12,10);
				\draw (0,9) -- (12,9);
				\draw (0,8) -- (12,8);
				\draw (0,7) -- (12,7);
				\draw (0,6) -- (12,6);
				\draw (0,5) -- (12,5);
				\draw (0,4) -- (12,4);
				\draw (0,3)-- (12,3);
				\draw (0,2) -- (12,2);
				\draw (0,1) -- (12,1);
				
				\node at (6,13.5) {\LARGE  \textbf{Target Vehicle Velocity}};
				\node[rotate = 90] at (-1.5,6) {\LARGE  \textbf{Ego Vehicle Velocity}};
				\node at (0.5,12.5) {\Large 5};
				\node at (1.5,12.5) {\Large  10};
				\node at (2.5,12.5) {\Large  15};
				\node at (3.5,12.5) {\Large  20};
				\node at (4.5,12.5) {\Large  25};
				\node at (5.5,12.5) {\Large  30};
				\node at (6.5,12.5) {\Large  35};
				\node at (7.5,12.5) {\Large  40};
				\node at (8.5,12.5) {\Large  45};
				\node at (9.5,12.5) {\Large  50};
				\node at (10.5,12.5) {\Large  55};
				\node at (11.5,12.5) {\Large  60};
				
				\node[rotate = 90] at (-0.5,11.5) {\Large  5};
				\node[rotate = 90] at (-0.5,10.5) {\Large  10};
				\node[rotate = 90] at (-0.5,9.5) {\Large  15};
				\node[rotate = 90] at (-0.5,8.5) {\Large  20};
				\node[rotate = 90] at (-0.5,7.5) {\Large  25};
				\node[rotate = 90] at (-0.5,6.5) {\Large  30};
				\node[rotate = 90] at (-0.5,5.5) {\Large  35};
				\node[rotate = 90] at (-0.5,4.5) {\Large  40};
				\node[rotate = 90] at (-0.5,3.5) {\Large  45};
				\node[rotate = 90] at (-0.5,2.5) {\Large  50};
				\node[rotate = 90] at (-0.5,1.5) {\Large  55};
				\node[rotate = 90] at (-0.5,0.5) {\Large  60};
		\end{tikzpicture}}
		\label{fig:obsfield}}
	\caption{Description of Obstructed Field Case}
	\label{descObF}
\end{figure}

\begin{figure}[h]
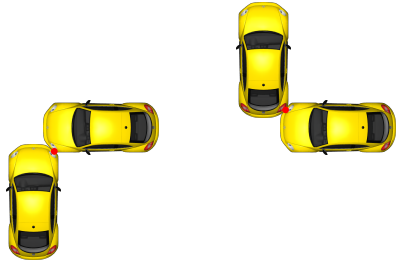

	\begin{center}
		\scalebox{0.9}{
			\begin{tikzpicture}
				
				\node at (-3.5,2){};
				\node at (0,2) {};
				
				\node at (-2.5,2) {\includegraphics[scale=0.15]{img/vwtop.png}};
				\node[rotate=-90] at (-3.5,0.9) {\includegraphics[scale=0.15]{img/vwtop.png}};
				
				\node at (1,2) {\includegraphics[scale=0.15]{img/vwtop.png}};
				\node[rotate=-90] at (-0.10,3) {\includegraphics[scale=0.15]{img/vwtop.png}};
				\node[circle, inner sep = 1pt, fill = red] at (-3.1971,1.6533) {};
				\node[circle, inner sep = 1pt, fill = red] at (0.2195,2.2672) {};
		\end{tikzpicture}}
	\end{center}	
	\caption[Different Collision Configurations]{Different Collision Configurations}
	\label{fig:config}
\end{figure}

Figure~\ref{fig:ttcradar} illustrates the setup for the calculation of the TTC, based on an a global X-Y reference frame. The equations can be easily adjusted to take into consideration that the radar gives information relative to the ego vehicle reference frame. The point at which the intersection can be given by the following expressions, given by Equations~\labelcref{eq:5.7,eq:5.8}:
{\small 
	\begin{align}
		x_+ &= \dfrac{(y_{ego}-y_{target})-(x_{ego}\cdotp \tan \theta_{ego} - x_{target} \cdotp \tan \theta_{target} )}{\tan \theta_{target} - \tan \theta_{ego}} \label{eq:5.7}\\
		y_+ &= \dfrac{(x_{ego}-x_{target})-(y_{ego}\cdotp \cot \theta_{ego} - y_{target} \cdotp \cot \theta_{target} )}{\cot \theta_{target} - \cot \theta_{ego}} \label{eq:5.8}
\end{align}}
After finding the intersection point, the Time-to-Reach (TTR) values for each vehicle can be obtained by Equations~\labelcref{eq:5.9,eq:5.10}:
{\small 
	\begin{align}
		\text{TTR}_{target} &= \dfrac{\sqrt{(x_+ - x_{target})^2 + (y_+ - y_{target})^2 }}{v_{target}} \label{eq:5.9}\\
		\text{TTR}_{ego} &= \dfrac{\sqrt{(x_+ - x_{ego})^2 + (y_+ - y_{ego})^2 }}{v_{ego}} \label{eq:5.10}
\end{align}}
To consider a more realistic case, it cannot be said that the vehicles will crash only when $\text{TTR}_{target} = \text{TTR}_{ego}$. Hence, it could be said that $\delta$ is a safety parameter described by Equation~\labelcref{eq:5.11}, such that:
{\small 
	\begin{equation}
		\label{eq:5.11}
		\text{TTC} = 
		\begin{cases}
			\min(\text{TTR}_{target},\text{TTR}_{ego}) & \text{if}\ |\text{TTR}_{target} - \text{TTR}_{ego}| \leq \delta \\
			\text{NaN} & \text{else}
		\end{cases}
\end{equation}}
To achieve similar performance between the simulated system and the reference real one provided by our partners, an expression of $\delta$ as a function of speed was reached by tuning the system to achieve the results described in Figure~\ref{fig:openfield} and Figure~\ref{fig:obsfield}. In the latter figures, the green square means that the collision was avoided using the radar AEB logic, and the red square means that a collision occurred. The tuning of $\delta$ was done taking into consideration speeds of vehicles that yield the following collision configurations during the braking process, as seen in Figure~\ref{fig:config}.

It is important to note that the AEB logic is a temporal one. Hence, it required a FSM finite state machine implemented by means of Stateflow on MATLAB. 

The acceleration controller block is PI controller that provides as an output a value of the required brake pedal (between $0$ and $1$) to achieve a certain deceleration value. It is important to note that the acceleration controller module is activated only when the AEB is taking place. The development of this ACC is based on work presented in \cite{ipgCM}. \textit{It is important to note that for our work, it is required that the driver can no longer press the gas when the AEB takes into action}.
\begin{figure}[t]
	\centering
	\scalebox{0.575}{
		\begin{tikzpicture}
			\fill[blue!30,opacity =0.5]  (-0.375,4.75) rectangle (8.625,-1.625);
			\draw[fill=green,draw = none]  (-5,4) rectangle (-3,3) node[pos=0.5, text width = 2cm,align = center]{\footnotesize Desired \\ Acceleration};
			\draw[fill=red,draw=none] (-5,2.5) rectangle (-3,1.5) node[pos=0.5, text width = 2cm,align = center]{\footnotesize Current\\ Acceleration};
			\draw [thick]   (-2,3.5) rectangle (-1.5,2);
			\node at (-1.75,3) {$+$};
			\node at (-1.75,2.5) {$-$};
			\draw[thick,-stealth] (-3,3.5) -- (-2.5,3.5) -- (-2.5,3) -- (-2,3);
			\draw[thick,-stealth] (-3,2)-- (-2.5,2) -- (-2.5,2.5) -- (-2,2.5);
			\draw[thick]  (-1.5,2.75) -- (-0.25,2.75) node [pos=0.5,above] (v3) {$\Delta a_x$};
			\draw [thick] (0.5,4.5) node (v1) {} -- (0.5,3.5) -- (1.5,4) node (v7) {} -- (0.5,4.5);
			\draw [thick] (0.5,2) node (v2) {} -- (0.5,1)-- (1.5,1.5) node (v8) {} -- (0.5,2);
			\node at (0.95,4) {\footnotesize $p_\text{gain}$};
			\node at (0.90,1.5) {\footnotesize $i_\text{gain}$};
			\draw[thick,-stealth] (-0.25,2.75) node (v4) {} -- (-0.25,4) -- (0.5,4);
			\draw[thick,-stealth] (-0.25,2.75) -- (-0.25,1.5) node (v5) {} -- (0.5,1.5);
			\draw [thick] (6,2) rectangle (7,1);
			\node at (6.5,1.5) {\Large $\Sigma$};
			\draw[thick,-stealth] (1.5,4) -- (5.5,4) -- (5.5,1.75) -- (6,1.75);
			\draw[thick,-stealth] (2,1.5) node (v6) {} -- (2,-0.5) --  (2.5,-0.5) node (v10) {};
			\draw[thick,-stealth] (1.5,1.5) -- (6,1.5);
			\draw [thick] (2.5,0) rectangle (3.5,-1);
			\draw[thick]  (4,0) rectangle (5,-1);
			\draw[thick,-stealth] (5,-0.5) -- (5.5,-0.5) -- (5.5,1.25) -- (6,1.25);
			\draw[thick,-stealth] (3.5,-0.5) node (v9) {} -- (4,-0.5);
			\draw[] (3,0) -- (3,-1);
			\draw[] (3.5,-0.5) -- (2.5,-0.5);
			\draw[] (3.5,-0.25) -- (3.25,-0.25) -- (3,-0.5) -- (2.75,-0.75) -- (2.5,-0.75);
			\draw[-stealth] (4.5,-0.75) -- (4.75,-0.75) -- (4.75,-0.25) -- (4.25,-0.25) -- (4.25,-0.75);
			\draw [thick]   (7.5,2) rectangle (8.5,1);
			\draw[thick,-stealth]  (7,1.5) -- (7.5,1.5) node (v11) {};
			\draw[] (8,2) -- (8,1);
			\draw[thick,-stealth] (8.5,1.5) node (v12) {} -- (9,1.5);
			\draw[]  (8.5,1.75) -- (8.25,1.75) -- (8,1.5) -- (7.75,1.25) -- (7.5,1.25);
			\draw[thick]  (7.5,1.5) -- (8.5,1.5);
			\node[right, text width = 2cm, align = center] at (8.375,1.375) {\footnotesize $u(k)$ \\ Brake Value};
			\node at (4.5,-1.25) {\footnotesize Memory};
			\node at (3,-1.25) {\footnotesize Saturation};
			\node[blue] at (4.125,5) {\footnotesize PI Controller};
	\end{tikzpicture}}
	\caption[Acceleration Control Architecture]{Acceleration Control Architecture}
	\label{fig:archacc}
\end{figure}
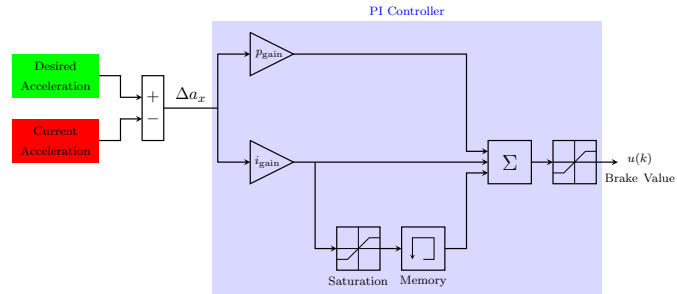

Figure~\ref{fig:archacc} shows the longitudinal acceleration controller. The architecture of the controller can be described as: Discrete PI Controller with Integral Anti-Windup. The control signal is calculated using the backward Euler discretization method~\cite{lee1992ieee} (where the gains were tuned), as given by Equations~\labelcref{eq:5.30,eq:5.321}:
\begin{align}
	&u(k) = \text{sat}(p_\text{gain}\Delta a_x +\text{sat}(i_\text{gain}\frac{T_s z}{z-1}\Delta a_x, A,B),A,B) \label{eq:5.30}\\
	&\text{sat}(x,A,B) = \min (\max(x,A),B) \label{eq:5.321}
\end{align}
Where:
\begin{conditions}
	p_{\text{gain}} & 0.001\\
	i_{\text{gain}} & 0.001\\
	Ts & 0.001 s\\
	A & Lower Limit of Saturation (1) \\
	B & Upper Limit of Saturation (0) 
\end{conditions}
\subsubsection{Connected System Model: Type A}
The connected system architecture shown in Figure \ref{fig:modifiedArch} was reached by making modifications to the previously developed Simulink model to take into consideration information from GPS, 5G, and Pirelli Cyber Tyre. It is important to note that the activation criteria remains the same for both the radar and 5G systems, as it is a characteristic of the type of collision the system is aimed at avoiding. 

Concerning 5G communications characteristics, literature suggests that the maximum value of delay accepted for this kind of applications is $10$ ms. This guarantees the minimum safety requirements~\cite{instance12x90}. The exact characteristics of the 5G network provided by Vodafone were used in the development; however, they could not be mentioned here for confidentiality purposes. 

Concerning the GPS information, CarMaker allows the modeling of the following errors~\cite{ipgCM,ipgCM2}: 
\begin{itemize}	
	\item \textit{Receiver Clock Error}: Arises because the receiver clock is not always synchronized with that of satellites. Standard deviation $= 5$ m with Correlation time $= 3600$ ms. 
	\item \textit{Ephemeris Error or Orbit Error}: Arises because of the small variations of the orbits of travel of the GPS satellites. Standard deviation $= 3$ m with Correlation time $= 1800$ ms. 
	\item \textit{Ionospheric Delay}: Arises because of ions in the ionosphere. Standard deviation  $= 5$ m with Correlation time $= 3600$ ms. 
	\item \textit{Tropospheric Delay}: Arises because of the refractions of the signal due to the variations of atmospheric conditons. Standard deviation $= 2$ m with Correlation time $= 1800$ ms. 
	\item \textit{Receiver Noise}: Pseudorange $= 0.1$ m, and Rated range = $0.05$ m 
\end{itemize}

Cocerning the Pirelli Cyber Tyre, it was mentioned earlier that it is able to provide information about the potential friction \textit{after} braking takes place. Hence, the use of this information is not optimal for the commercial braking logic, as it is a static velocity-dependent logic. For example, even if the potential ground friction was measured to be low while braking is taking place, nothing can be done about it. Considering the current AEB logic, it is clearly seen that the nature of the logic prohibits utilizing the Ego Vehicle's Pirelli Cyber Tyre to its full potential. 

\begin{figure}[h]
	\centering
	\scalebox{0.45}{
		\begin{tikzpicture}
			
			\draw[fill =blue!30]  (-1,2) rectangle (2.5,0) node [pos = 0.5, text width = 3 cm, align = center] {Velocity-dependent Braking Logic};

			\draw[thick]  (-1.125,0.5) ellipse (0.125 and 0.125);
			\draw[thick]   (-1.125,1.5) ellipse (0.125 and 0.125);
			\draw[thick]  (-1.125,0.625) -- (-1.625,1.5);
			\draw[thick,stealth-stealth] (-1.6995,1.1071) arc (139.9978:100:0.75);

			\draw[thick,-stealth] (2.5,1) -- (4,1)node [pos = 0.57, above] {dec($v$)};
			\draw[fill = red!30]  (6,2) rectangle (9.5,0) node [pos = 0.5, text width = 3 cm, align = center] {Acceleration Controller};
			\draw[thick,-stealth] (7.75,0) -- (7.75,-1.5) ;
			\node at (7.75,-2.5) {\includegraphics[scale=0.28]{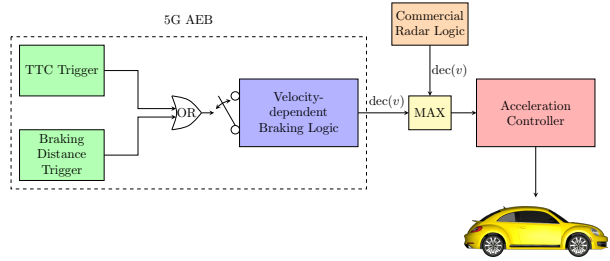}};
			
			\draw[thick]  plot[smooth, tension=.7] coordinates {(-3,1.5) (-2.5,1.375) (-2.125,1)};
			\draw[thick]  plot[smooth, tension=.7] coordinates {(-3,0.5) (-2.5,0.625) (-2.125,1)};
			\draw[thick]  plot[smooth, tension=.7] coordinates {(-3,1.5) (-2.875,1.125) (-2.875,0.875) (-3,0.5)};
			\node at (-2.5625,1) {OR};
			\draw[thick,-stealth] (-2.125,1) -- (-1.75,1);
			
			\draw[fill = green!30]  (-7.5,3) rectangle (-5,1.5) node [pos = 0.5]{TTC Trigger};
			\draw[fill = green!30]  (-7.5,0.5) rectangle (-5,-1)node [pos = 0.5, text width = 3cm, align = center]{Braking \\ Distance \\ Trigger};
			\draw[thick,-stealth] (-5,2.25) -- (-4,2.25) -- (-4,1.125) -- (-2.875,1.125);
			
			\draw[thick,-stealth] (-5,-0.25) -- (-4,-0.25) -- (-4,0.875) -- (-2.875,0.875);
			\draw [fill=yellow!30] (4,1.5) rectangle (5.25,0.5)node [pos= 0.5]{MAX};
			\draw[thick,-stealth] (5.25,1) -- (6,1);
			\draw[thick,-stealth] (4.625,3) -- (4.625,1.5);
			\node[right] at (4.55,2.25) {dec($v$)};
			\draw[fill = orange!30]  (3.5,4.25) rectangle (5.75,3) node [pos =0.5, text width = 3cm,align = center]{Commercial \\ Radar  Logic};
			\draw[dashed]  (-7.75,3.25) rectangle (2.75,-1.25);
			\node at (-2.5,3.75) {5G AEB};
	\end{tikzpicture}}
	\caption[Scheme of Upgraded System]{Detailed Scheme of Upgraded System A}
	\label{fig:kkd2}
\end{figure}

As an initial attempt to utilize information about potential ground friction, it is assumed that it is available from an active map. This assumption is realistic, for example, in a scenario where several vehicles are driving in the same area and sharing their potential ground friction estimation that is so collected in an active map available to the ego vehicle. For this reason, it was reasonable to include braking distance in the logic. The scheme is shown in Figure~\ref{fig:kkd2}. Now, the trigger to the system is no longer the TTC only, but also the braking distance. If the braking distance becomes equal to a certain threshold related to the distance to the collision point, the braking maneuver takes place.

\begin{figure}[h]
	\centering
	\scalebox{0.5}{		
		\begin{tikzpicture}
			\node at (10,-3) {$0$};
			\node at (20.5,-3)  {$v_\text{max}$};
			\node at (17,-3)  {$v_\text{a}$};
			\node at (13.5,-3) {$v_\text{b}$};
			\draw[fill = green!30,draw = none] (20.5,-1) rectangle (17,-2.5) ;
			\draw[fill = green!30,draw = none]   (17,-1) rectangle (15.5,-2.5) node (v1) {};
			\draw[fill = blue!30,opacity = 0.3,draw = none]  (15.5,-2.5) rectangle (13.5,0);
			\draw[fill =red!30,opacity = 0.3,draw = none] (13.5,-2.5) node (v3) {} rectangle (10,0.5);
			\draw[fill=red!30, draw = none]  (10,-0.5) rectangle (13.5,-2.5);
			\draw[fill = blue!30,draw = none]  (13.5,-0.5) rectangle (15.5,-2.5);
			\draw[thick] (10,-2.25) -- (10,-2.75);
			\draw[thick] (20.5,-2.25) -- (20.5,-2.75);
			\draw[thick] (13.5,0.5) -- (13.5,-2.75);
			\draw[thick](17,0.5) -- (17,-2.75);
			
			\draw[thick,-stealth] (9.75,-2.5) -- (9.75,2);
			\node[rotate = 90] at (9.5,-0.25) {$|\text{deceleration}|$};
			\node[opacity = 0.3] at (11.625,0.75) {Level C};
			\node[opacity = 0.3] at (14.5,0.25) {Level B};
			\node at (16.25,-0.75) {Level A};
			\node at (17.75,-0.75) {Level A};
			\node at (15.25,-2.75) {Stage B};
			\node at  (18.75,-2.75) {Stage A};
			\node at (11.75,-2.75) {Stage C};
			\draw[thick,dashed] (15.5,0) -- (15.5,-2.5);
			\draw[thick,stealth-stealth] (15.5,-1.75) -- (17,-1.75) node [pos = 0.5,above]{300 ms};
			\draw[ultra thick] (20.5,-0.5) -- (10,-0.5) node (v2) {};
			
			\node at (11.625,-0.25) {Level C};
			\node at (14.5,-0.25) {Level B};
			
			\node at (-5,-3) {$0$};
			\node at (5.5,-3)  {$v_\text{max}$};
			\node at (2,-3)  {$v_\text{a}$};
			\node at (-1.5,-3) {$v_\text{b}$};
			\draw[fill = green!30,draw = none] (5.5,-1) rectangle (2,-2.5) ;
			\draw[fill = green!30,draw = none]   (2,-1) rectangle (0.5,-2.5) node (v1) {};
			\draw[fill = blue!30,draw = none]  (0.5,-2.5) rectangle (-1.5,0);
			\draw[fill =red!30,draw = none] (-1.5,-2.5) rectangle (-5,0.5);

			\draw[thick] (-5,-2.25) -- (-5,-2.75);
			\draw[thick] (5.5,-2.25) -- (5.5,-2.75);
			\draw[thick] (-1.5,0.5) -- (-1.5,-2.75);
			\draw[thick](2,0.5) -- (2,-2.75);
			
			\draw[thick,-stealth] (-5.25,-2.5) -- (-5.25,2);
			\node[rotate = 90] at (-5.5,-0.25) {$|\text{deceleration}|$};
			\node at (-3.375,0.75) {Level C};
			\node at (-0.5,0.25) {Level B};
			\node at (1.25,-0.75) {Level A};
			\node at (3.75,-0.75) {Level A};
			\node at (0.25,-2.75) {Stage B};
			\node at  (3.75,-2.75) {Stage A};
			\node at (-3.25,-2.75) {Stage C};
			\draw[thick,dashed] (0.5,0) -- (0.5,-2.5);
			\draw[thick,stealth-stealth] (0.5,-1.75) -- (2,-1.75) node [pos = 0.5,above]{300 ms};
			
			\draw[thick,-stealth]  plot[smooth, tension=.7] coordinates {(5.6,-0.6) (7.4,0) (9.2,-0.6)};
			\node[right, text width = 3.5cm] at (5.75,-1.8) {\small Based on $\mu$, the maximum possible deceleration achievable for each stage is modified  };
			
	\end{tikzpicture}}
	\caption{Modification of Braking Scheme}
	\label{descOfxf}
\end{figure}
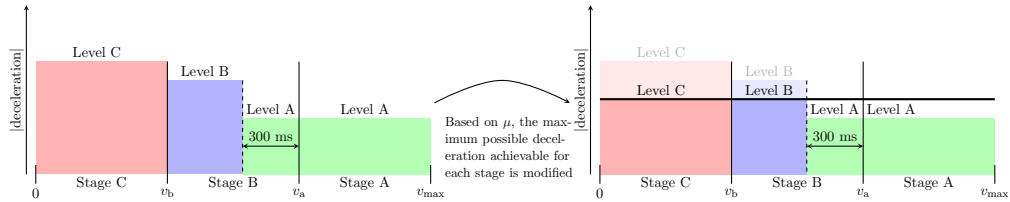

It is noteworthy that the parameters of the braking logic weren't modified: this first novel approach assumed that the ECU containing the AEB logic can't be modified. In fact, this approach can be considered as an add-on with respect to the commercial configuration with the minimum modification required. The braking distance is calculated based on a the braking scheme originally described in Figure~\ref{fig:scheme}. The maximum deceleration level was modified based on the potential ground friction measurement, as described in Figure~\ref{descOfxf}, and then the braking distance was calculated by performing numerical integration.

It could be seen that the scheme provided in Figure \ref{fig:kkd2} complies with the system architecture provided in Figure \ref{fig:modifiedArch}. Here, information coming from 5G about the position ad speed of nearby vehicles is used to evaluate TTC, and the BD based on the received potential ground friction. Once either one of the triggers is activated, a deceleration request is sent to the acceleration controller. 

\subsubsection{Connected System Model: Type B}

This approach means a deeper modification of the vehicle original hardware and software configuration in order to obtain better results in terms of safety. In an attempt for further development, it was decided to modify the braking logic block. This is because the current braking logic does not facilitate taking advantage of the potential of he information received from the Pirelli Cyber Tyre. It is important to explore the option of giving the vehicle a reference deceleration that is adaptable to the road conditions: for example: increases exponentially throughout the braking maneuver. This approach tries to mimic the commercial braking logic, in that the braking logic requests an initial deceleration value of around $30\%$ of the maximum possible deceleration, and reaches $80\%$ of its maximum value (this allows the driver to still have control on the vehicle at maximum braking). The formulation is given by \cref{eq:6.8}:
\begin{equation}
	\label{eq:6.8}
	a(t) = A e^{\alpha t}
\end{equation}
Where $	A = - \max \ (k_\text{min}\mu  g, a_\text{current})$. The velocity profile is given by \cref{eq:6.9}:
\begin{equation}
	\label{eq:6.9}
	\begin{aligned}
		v(t) &=v_0 + \int_{0}^{t} a(t) \text{d}t \\
		&= v_0 + \frac{A}{\alpha}e^{\alpha t} - \frac{A}{\alpha}
	\end{aligned}
\end{equation}
The displacement of the vehicle is given by \cref{eq:6.10}:
\begin{equation}
	\label{eq:6.10}
	\begin{aligned}
		s(t) &= \int_{0}^{t} v_0 + \frac{A}{\alpha}e^{\alpha t} - \frac{A}{\alpha} \ \text{d}t  \\
		& = \frac{A}{\alpha^2}e^{\alpha t} + Bt + C
	\end{aligned}
\end{equation}
Now, it is considered that the vehicle must reach a complete stop at time $T$ ($v(T) = 0$), such that $a(T) \leq -a_\text{max} \mu g$ and $s(t)\leq d$, where $d$ is the distance to the intersection. Due to complex nature of the problem, the following approach was considered, as seen in \cref{fig:sysd}:
\begin{figure}[h]
	\centering
	\scalebox{0.45}{
		\begin{tikzpicture}
			
			\draw[fill =blue!30]  (-1,2) rectangle (2.5,0) node [pos = 0.5, text width = 3 cm, align = center] {Exponential Deceleration Request};

			\draw[thick]  (-1.125,0.5) ellipse (0.125 and 0.125);
			\draw[thick]   (-1.125,1.5) ellipse (0.125 and 0.125);
			\draw[thick]  (-1.125,0.625) -- (-1.625,1.5);
			\draw[thick,stealth-stealth] (-1.6995,1.1071) arc (139.9978:100:0.75);

			\draw[thick,-stealth] (2.5,1) -- (4,1)node [pos = 0.57, above] {dec($v$)};
			\draw[fill = red!30]  (6,2) rectangle (9.5,0) node [pos = 0.5, text width = 3 cm, align = center] {Acceleration Controller};
			\draw[thick,-stealth] (7.75,0) -- (7.75,-1.5) ;
			\node at (7.75,-2.5) {\includegraphics[scale=0.3]{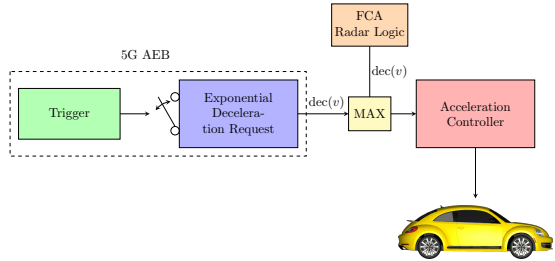}};

			\draw [fill=yellow!30] (4,1.5) rectangle (5.25,0.5)node [pos= 0.5]{MAX};
			\draw[thick,-stealth] (5.25,1) -- (6,1);
			\draw (4.625,3) -- (4.625,1.5);
			\node[right] at (4.55,2.25) {dec($v$)};
			\draw[fill = orange!30]  (3.5,4.25) rectangle (5.75,3) node [pos =0.5, text width = 3cm,align = center]{FCA \\ Radar  Logic};
			\node at (-2,2.75) {5G AEB};
			\draw[thick,-stealth] (-3.75,1) -- (-1.875,1);
			\draw[dashed] (-6,2.25) rectangle (2.75,-0.25);
			\draw[fill = green!30] (-5.75,1.75) rectangle (-2.75,0.25)node [pos = 0.5] {Trigger};
	\end{tikzpicture}}
	\caption[Detailed Scheme of Upgraded System B]{Detailed Scheme of Upgraded System B}
	\label{fig:sysd}	
\end{figure}

At every loop, the \textbf{trigger} block solves the following to get values of $\alpha$ and $T$:
\begin{equation*}
	v(T)= 0, \ s(T) = d
\end{equation*}
Once the values of $\alpha$ and $T$ are obtained, $a(T) = A e^{\alpha T}$ is calculated. If $a(T)\geq a_\text{max} \mu g - a_\text{tuned}$ (where $a_\text{tuned}$ is a certain threshold that is tuned, for our case it is $1 \ \text{m}/\text{s}^2$), then the trigger is activated and the braking maneuver is initiated. 

It could be immediately seen that the system still has an issue with the characteristic of the Pirelli Cyber Tyre (since it provides an estimate of the value of potential ground friction at the moment of braking). However, the advantage here is that the system is somehow adaptable, providing a higher deceleration value as a compensation. The adaptability of the system is highly affected by the value of $a_\text{tuned}$.
\section{Simulation Testing Results}
\subsection{Upgraded System Type A: Normal Ground Friction Environment}
\begin{figure}[h]
	\centering
	\subfloat[]{
		\includegraphics[scale=0.275]{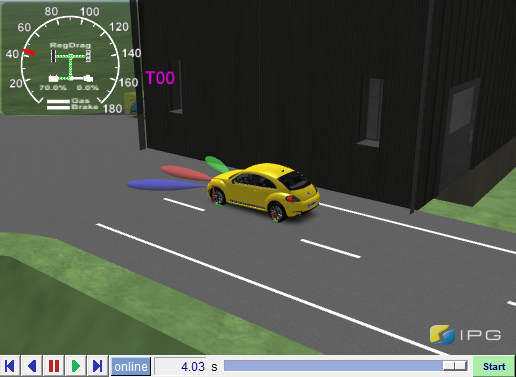}
		}
	\centering
	\subfloat[]{
		\includegraphics[scale=0.275]{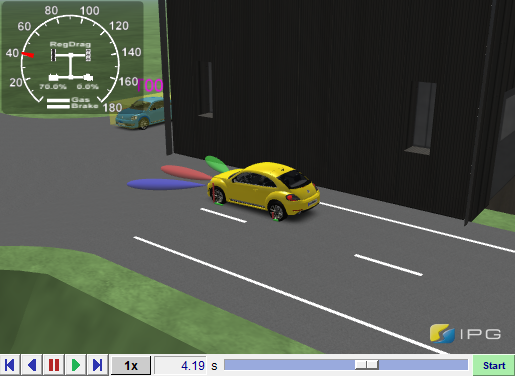}
		}
	\centering
	\subfloat[]{
		\includegraphics[scale=0.275]{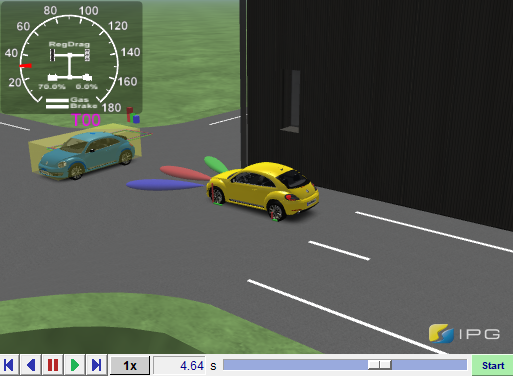}
	}
	\caption[]{Screenshots from CarMaker in Chronological Order: System Type A, Normal Ground Friction}
	\label{fig:bd}
\end{figure}

\begin{figure}[h]
	\centering
	\subfloat[]{
		\includegraphics[scale=0.3]{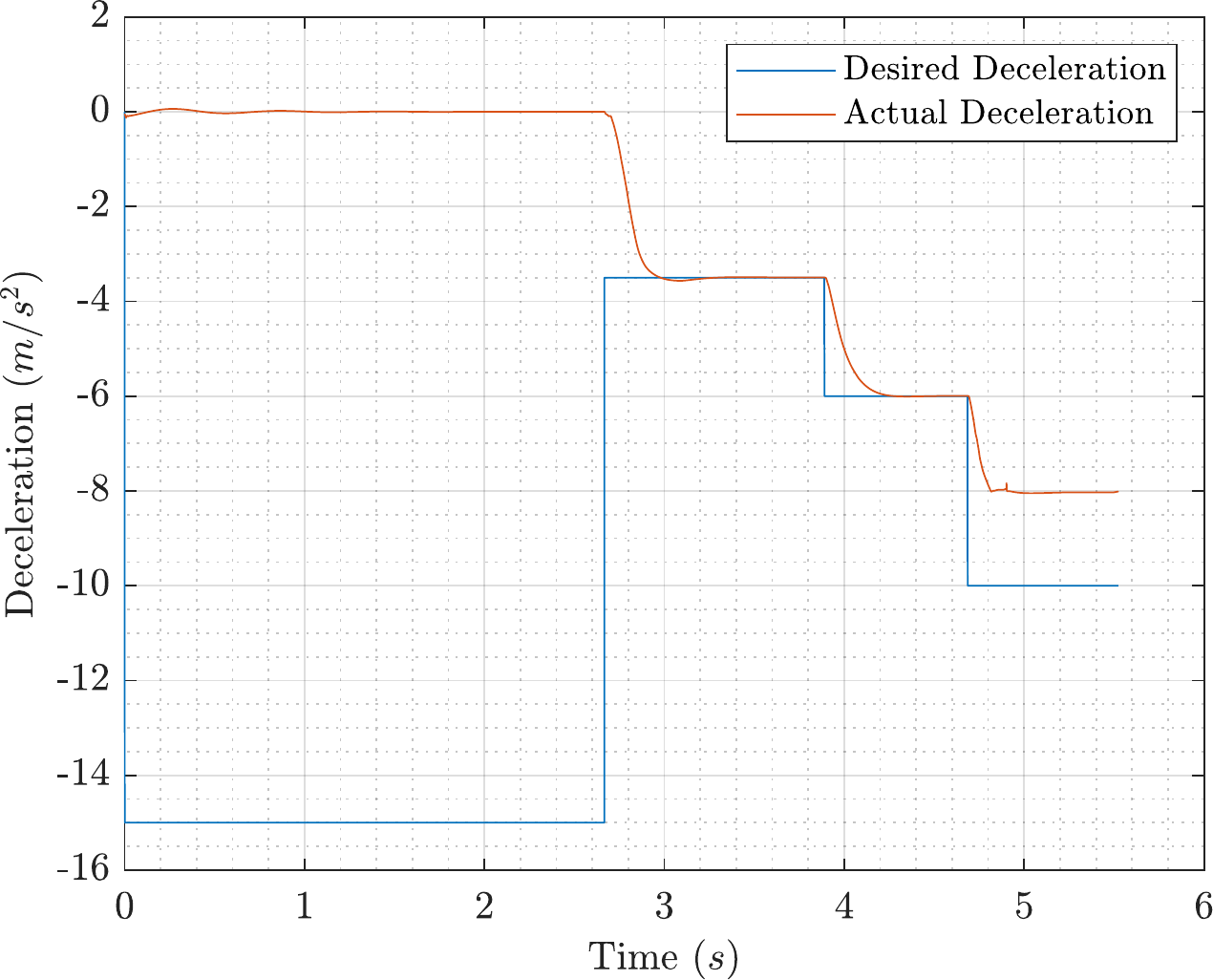}
		\label{fig:GBD1}}
	\subfloat[]{	
		\includegraphics[scale=0.3]{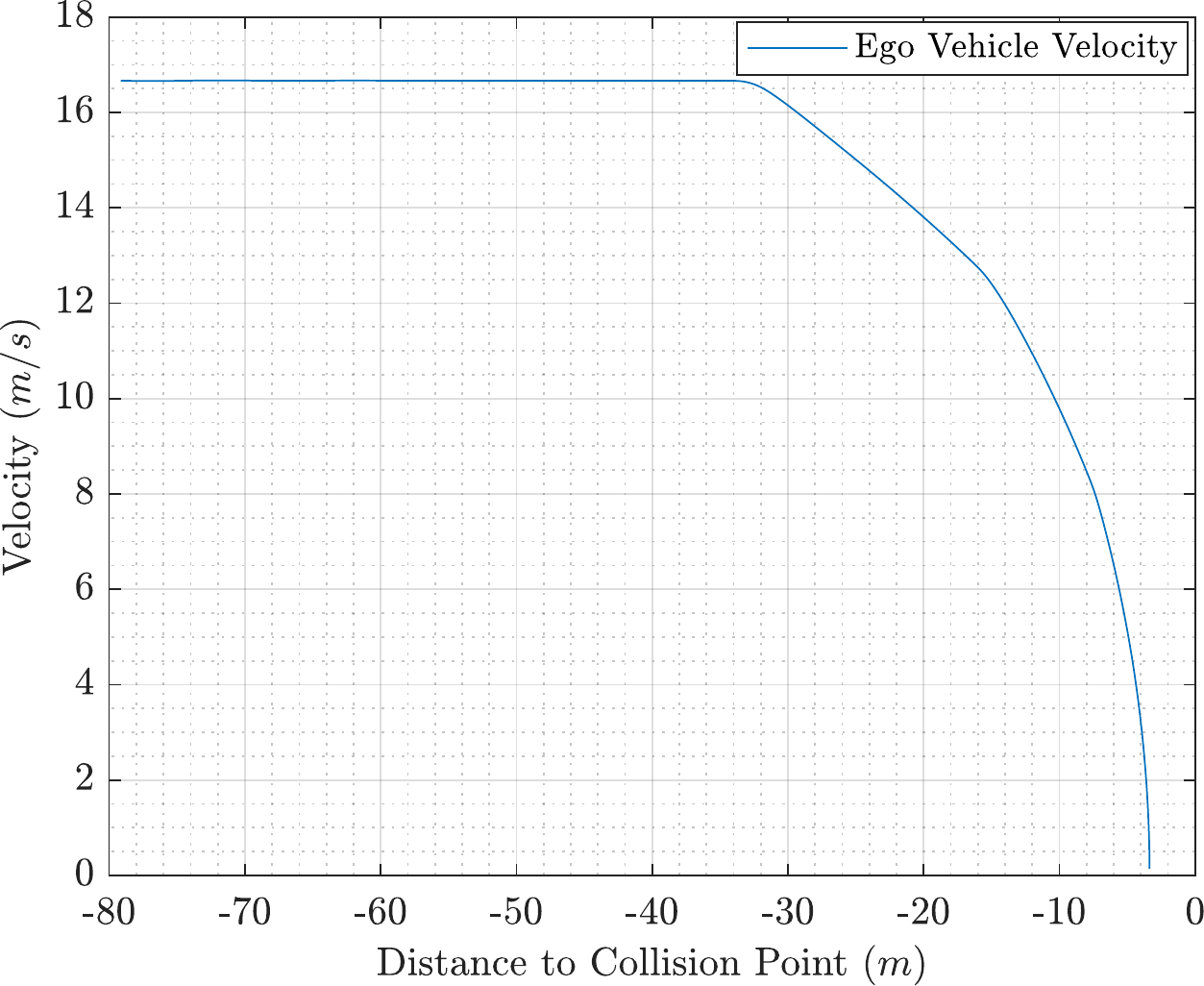}
		\label{fig:GBD2}}
	\caption{System Type A: Normal Ground Friction, (a) Desired Deceleration vs Actual Deceleration (b) Ego Vehicle Velocity vs Distance to Collision Point}
	\label{totall}
\end{figure}

The test was carried out with a ground friction coefficient equal to $0.85$ . The test results displayed are for the case when both vehicles approach the intersection traveling at 60 km/h. It is noteworthy that the system was tested for all velocity combinations as outlined in Figure~\ref{fig:obsfield}. However, as a choice of displaying results in this paper, the case of having both vehicles traveling at high velocities was chosen, as it is usually the case causing collision without connectivity as well as with low ground friction. The modified system was tested in a normal friction conditions, as seen in Figure \ref{fig:bd}. It could be clearly seen in \ref{fig:GBD1} that the maximum deceleration a vehicle can reach in normal ground friction conditions is approximately $8.3 \text{m/s$^2$}$. 

From Figure~\ref{fig:bd} it could be seen that the vehicle avoids the collision. Also, the vehicle approaches the intersection at a reduced speed, and the stopping takes place before the pre-defined collision point; this is quite important as the AEB action turns off when the vehicle reaches a complete stop.

Relevant numerical results are presented in Figure \ref{totall}.

Figure \ref{fig:GBD1} shows the deceleration request as well as the actual one. Note that the value of $-15$ $\text{m}/\text{s}^2$ is the equivalent of no deceleration request. It could be seen in Figures \ref{fig:GBD1} and \ref{fig:GBD2} that once the condition described in Equation \ref{eq:5.11} is satisfied, the AEB system is activated. 
\subsection{Upgraded System Type A: Low Ground Friction Environment}
The modified system was tested in a low friction condition ($\mu = 0.4$). 
\begin{figure}[h]
	\centering
\subfloat[]{
		\includegraphics[scale=0.275]{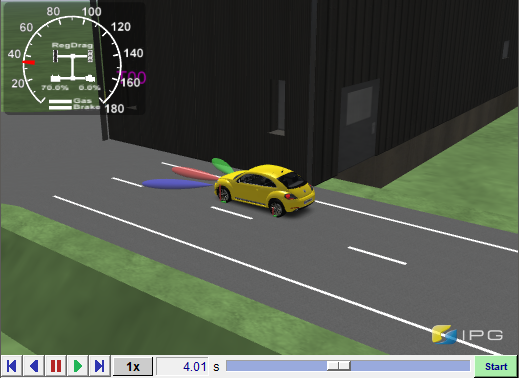}}
\subfloat[]{
		\includegraphics[scale=0.275]{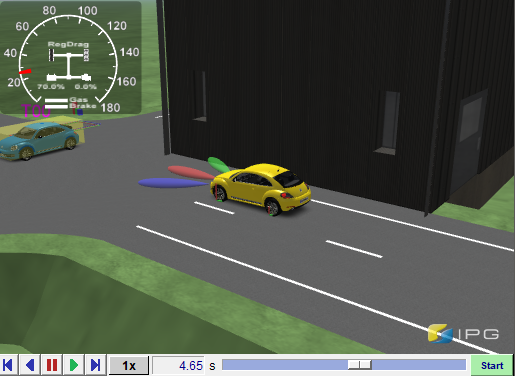}}
\subfloat[]{
		\includegraphics[scale=0.275]{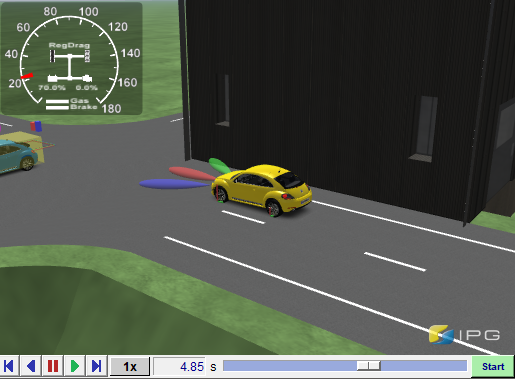}}
	\caption[Screenshots from CarMaker in Chronological Order]{Screenshots from CarMaker in Chronological Order: System Type A, Low Ground Friction}
	\label{fig:bdL}
\end{figure}

\begin{figure}[h]
	\centering
	\subfloat[]{
		\includegraphics[scale=0.3]{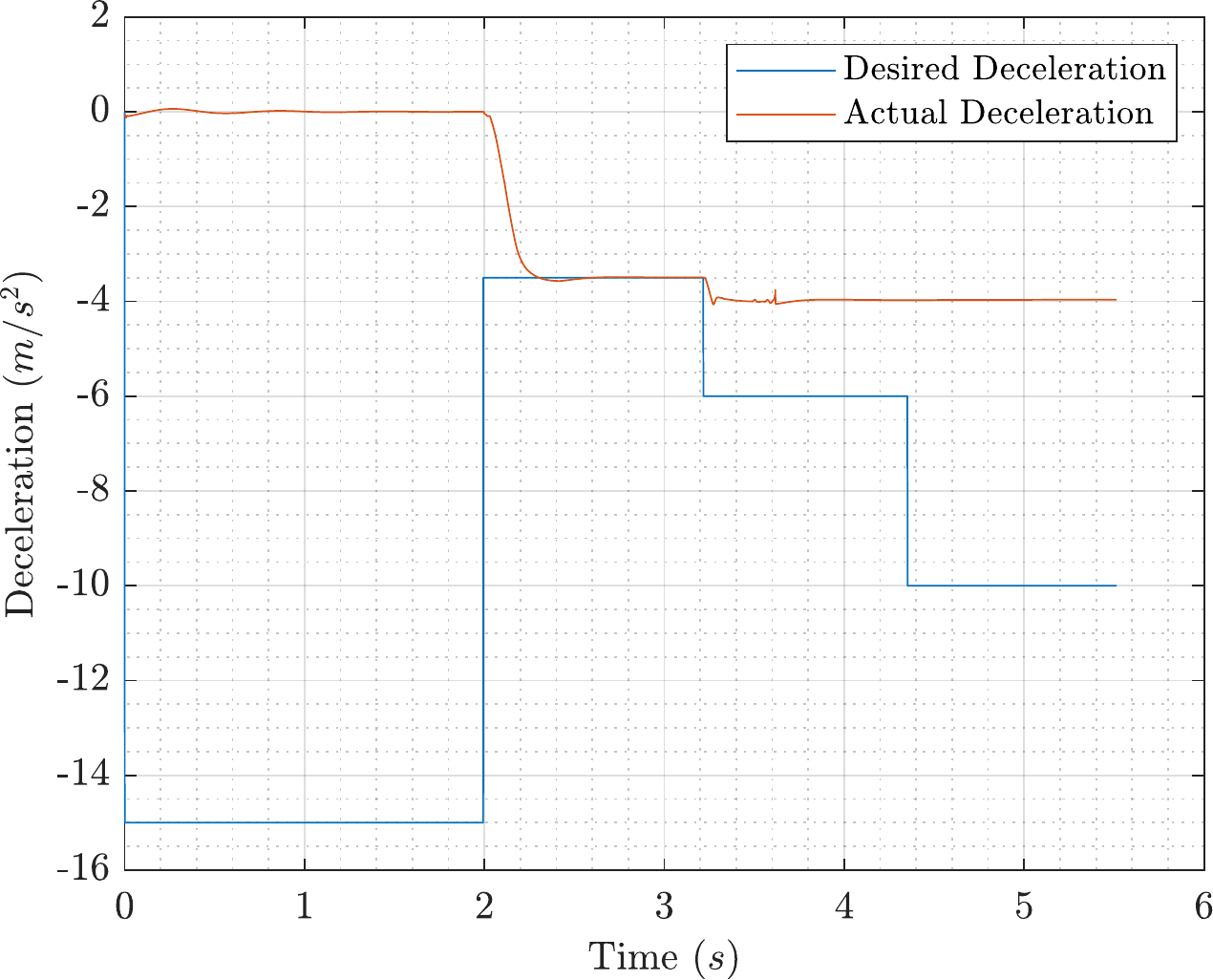}
		\label{fig:GBDl1}}
	\subfloat[]{	
		\includegraphics[scale=0.3]{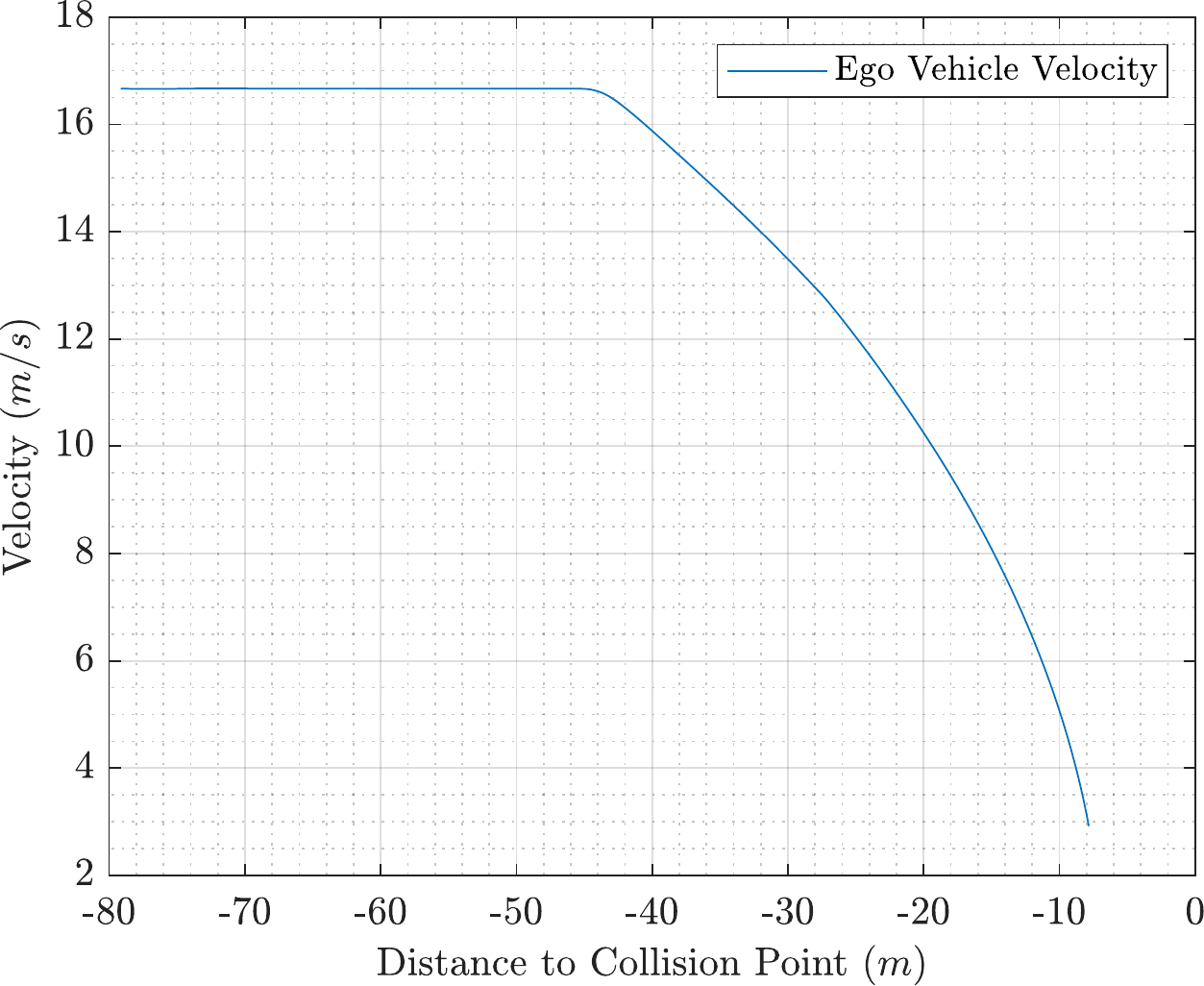}
		\label{fig:GBDl2}}
	\caption{System Type A: Low Ground Friction, (a) Desired Deceleration vs Actual Deceleration (b) Ego Vehicle Velocity vs Distance to Collision Point}
	\label{fig:xx}
\end{figure}

Relevant numerical results are presented in Figure \labelcref{fig:xx}. 

From Figure~\ref{fig:bdL} it could be seen that the vehicle avoids the collision. Also, the vehicle approaches the intersection at a reduced speed, and the stopping takes place before the pre-defined collision point. Furthermore, comparing Figures~\ref{fig:GBD2} and \ref{fig:GBDl2}, it could be seen that in low ground friction, the braking starts 34 meters and 44 meters before the predicted collision point, respectively. 

It could be seen in Figure \ref{fig:GBDl1} that the vehicle is unable to reach the deceleration requested as a result of the ground condition.

It is clear that the only way for the system to avoid the collision in this case is activating the AEB system earlier thorough the braking distance calculation which utilizes a priori knowledge about ground friction. The difference could be clearly seen when comparing Figure \ref{fig:GBD2} and Figure \ref{fig:GBDl2}. 

\subsection{Upgraded System Type B: Normal Ground Friction Environment}
\begin{figure}[h]
	\centering
\subfloat[]{
		\includegraphics[scale=0.275]{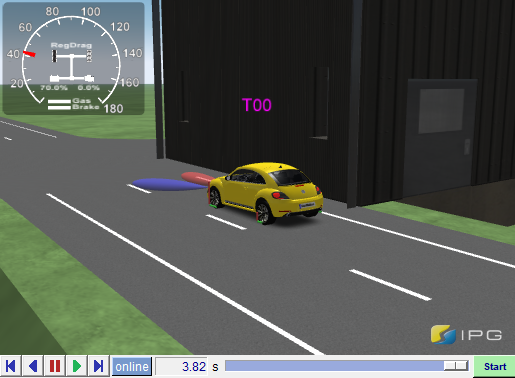}}
	\subfloat[]{
		\includegraphics[scale=0.275]{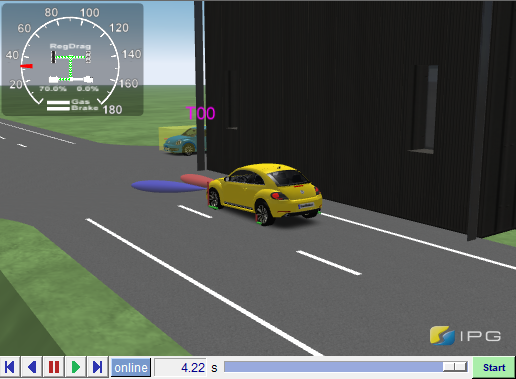}}
	\subfloat[]{
		\includegraphics[scale=0.275]{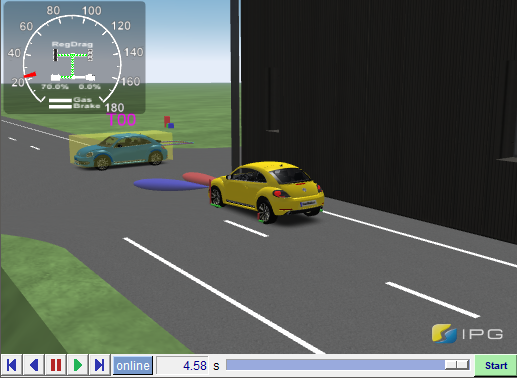}}
	\caption[Screenshots from CarMaker in Chronological Order]{Screenshots from CarMaker in Chronological Order: System Type B, Normal Ground Friction}
	\label{fig:bddD}
\end{figure}
\begin{figure}[h]
	\centering
	\subfloat[]{
		\includegraphics[scale=0.3]{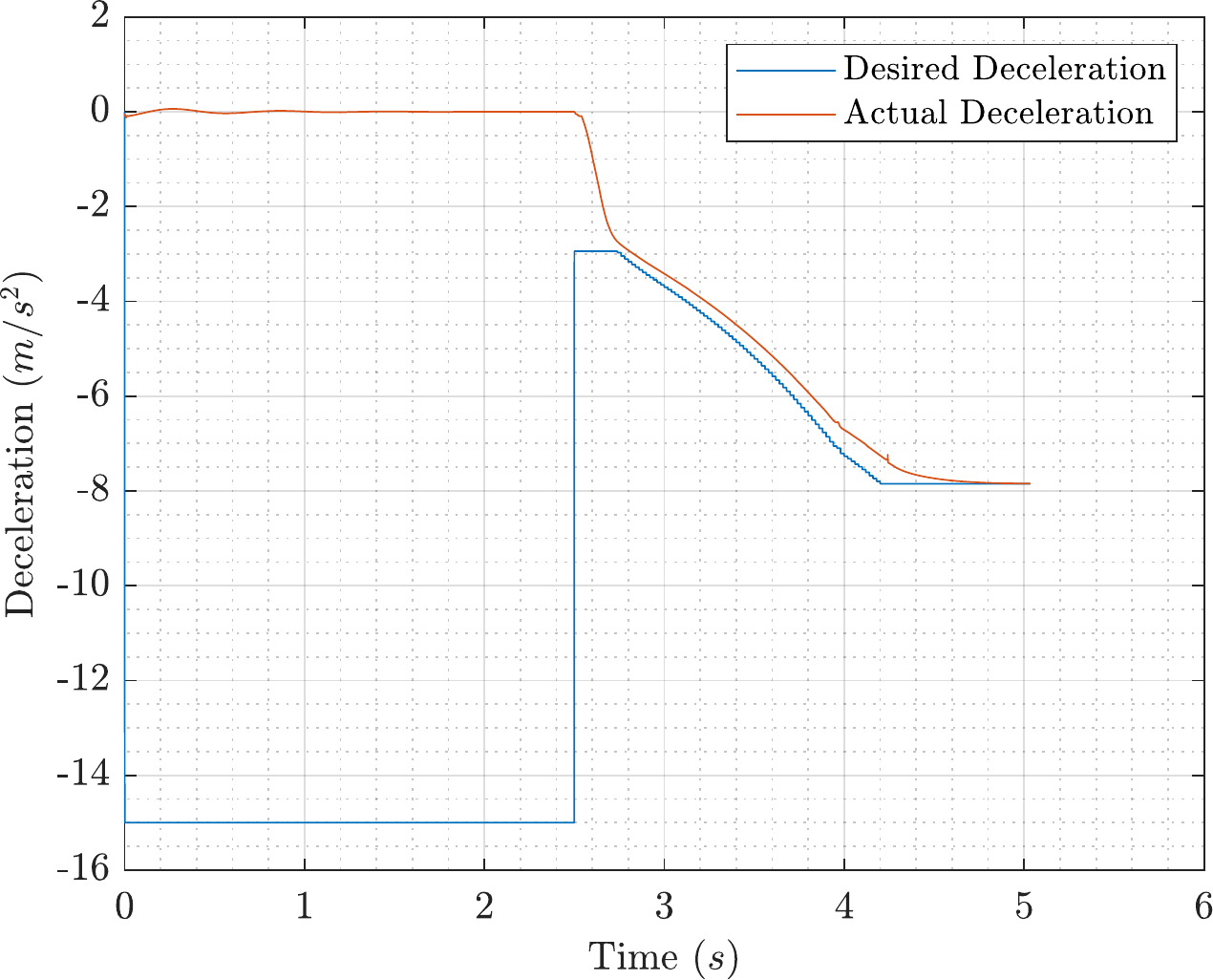}
		\label{fig:GD1}}
	\subfloat[]{	
		\includegraphics[scale=0.3]{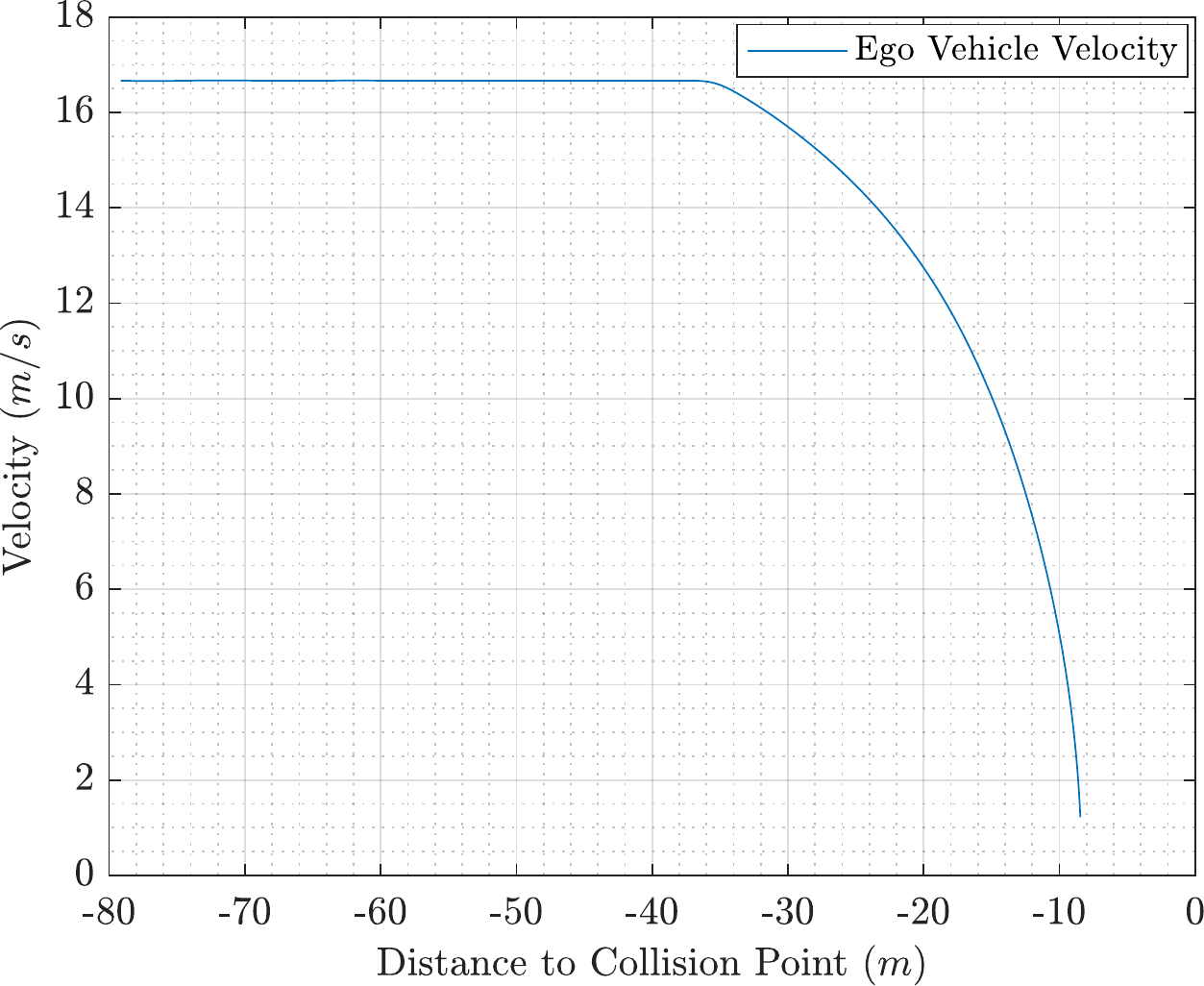}
		\label{fig:GD2}}\\
	\caption{System Type B: Normal Ground Friction, (a) Desired Deceleration vs Actual Deceleration (b) Ego Vehicle Velocity vs Distance to Collision Point }
	\label{xssld}
\end{figure}

Relevant numerical results are presented in Figure \ref{xssld}. It could be seen from Figure \ref{fig:GD1} that the braking maneuver follows an exponential deceleration scheme. Hence, with respect to the stepped scheme in System A, System B provides a smoother profile that reflects an improvement in driving  comfort.


When inspecting \cref{fig:bddD}, it could be seen that the vehicle successfully avoids the collision. Also, the vehicle stops completely before the collision point. 
\subsection{Upgraded System B: Low Ground Friction Environment}
\subsubsection{Friction Known A Priori}
It would be interesting to investigate the case where the low ground friction value ($\mu=0.4$) is known before the braking maneuver takes place.
\begin{figure}[h]
	\centering
\subfloat[]{
		\includegraphics[scale=0.275]{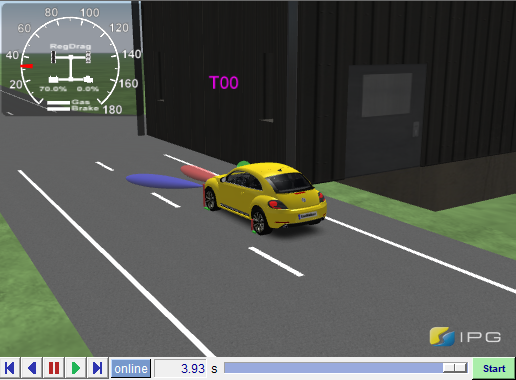}}
\subfloat[]{
		\includegraphics[scale=0.275]{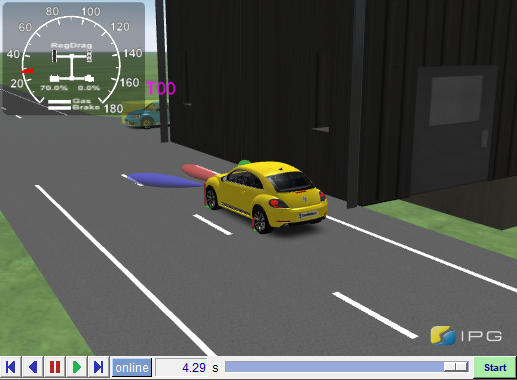}}
\subfloat[]{
		\includegraphics[scale=0.275]{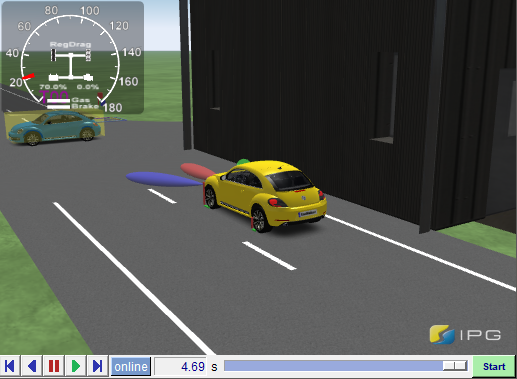}}

	\caption[Screenshots from CarMaker in Chronological Order]{Screenshots from CarMaker in Chronological Order: System Type B, Low Ground Friction}
	\label{fig:bdd}
\end{figure}

\begin{figure}[h]
	\centering
	\subfloat[]{
		\includegraphics[scale=0.3]{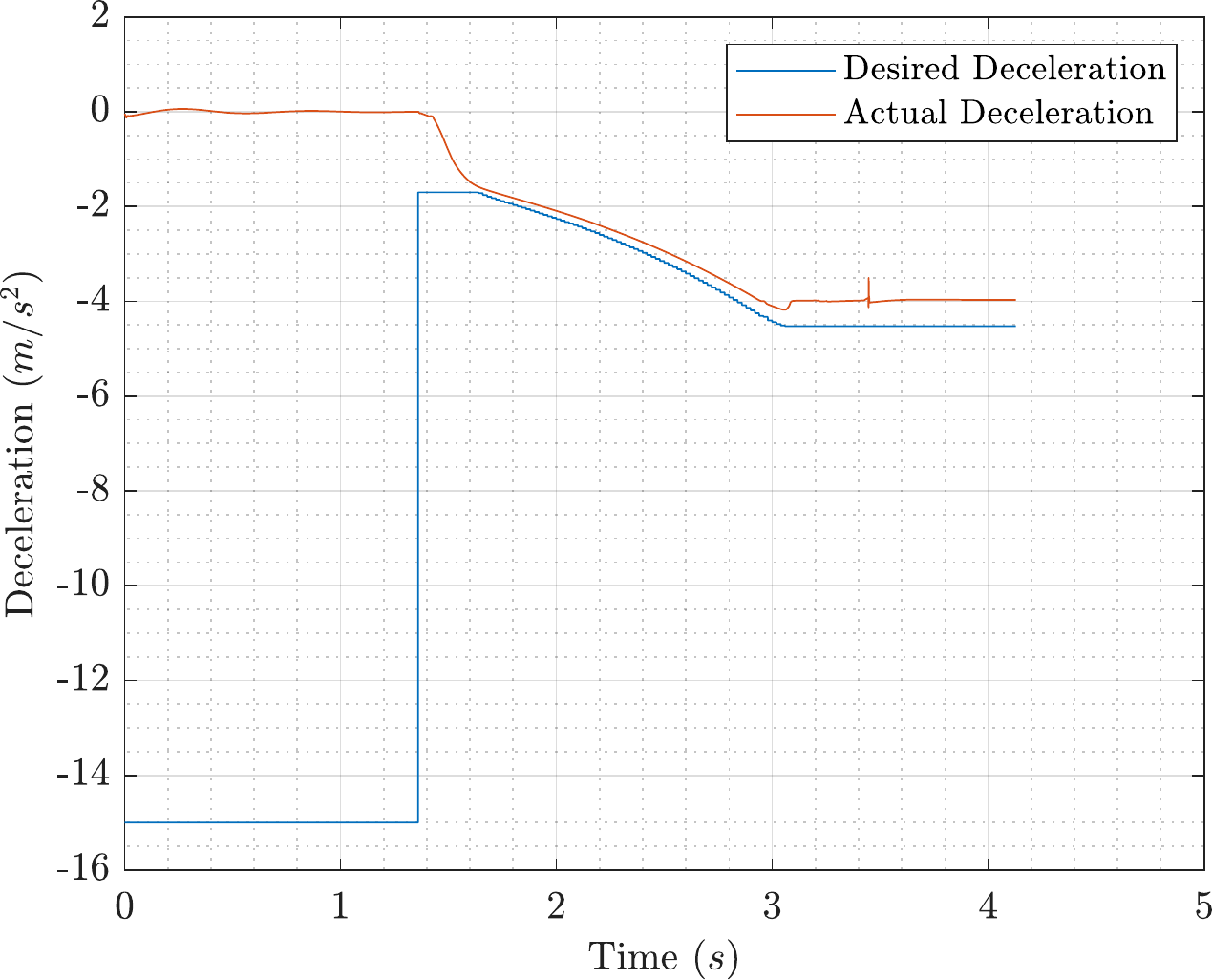}
		\label{fig:DLKF1}}
	\subfloat[]{	
		\includegraphics[scale=0.3]{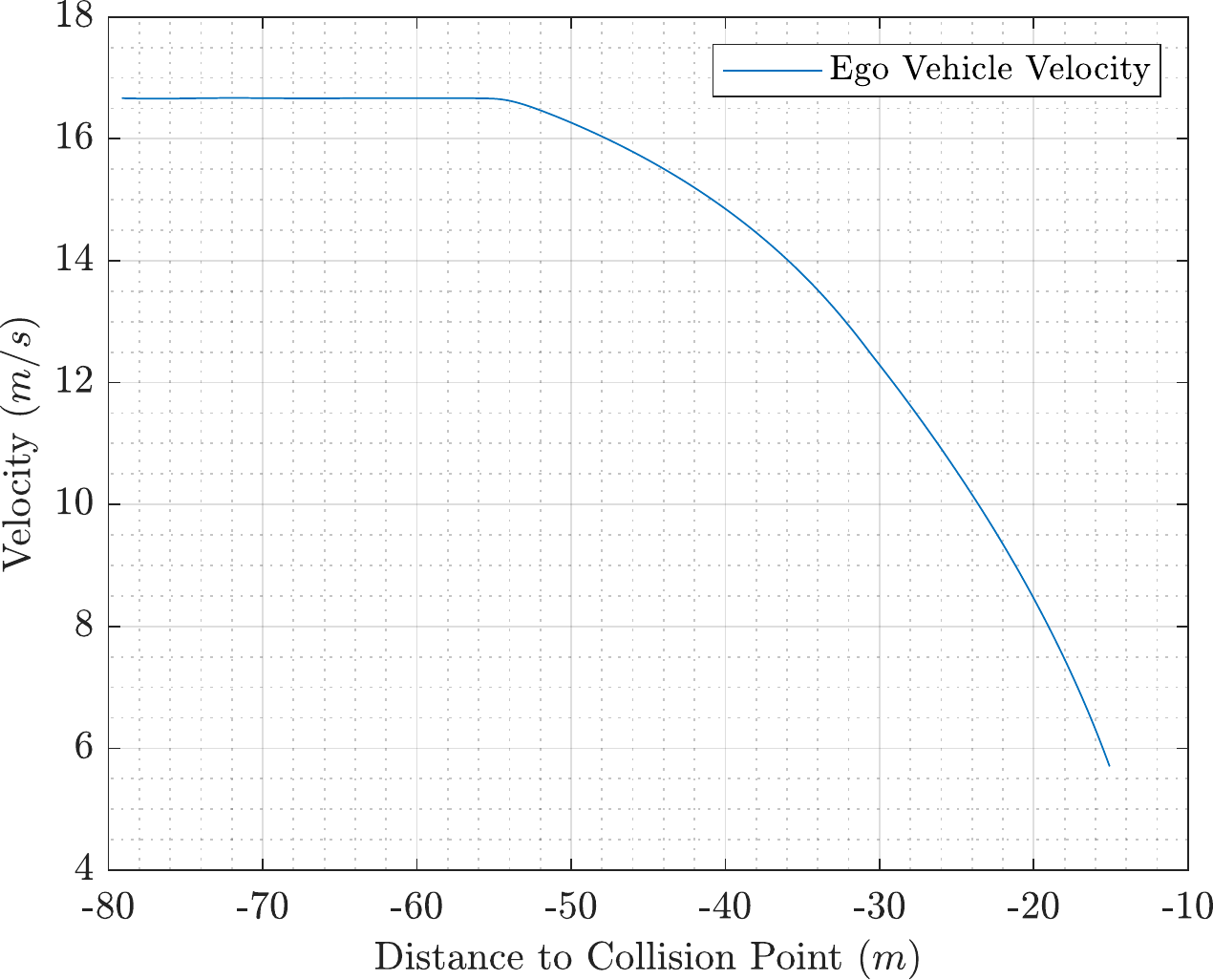}
		\label{fig:DLFK2}}\\
	\caption{System Type B: Low Ground Friction, (a) Desired Deceleration vs Actual Deceleration (b) Ego Vehicle Velocity vs Distance to Collision Point }
	\label{fig:xxxxx}
\end{figure}

Relevant numerical results are presented in Figure \ref{fig:xxxxx}. It could be seen that the vehicle successfully stops before the intersection, and avoids the collision. It could be seen from Figure \ref{fig:DLKF1} that the braking maneuver is smooth and comfortable. When comparing the performance of System B in low friction with that of System A (Figure \ref{fig:GBDl1}), it could be seen that the system manages to stop further away from the intersection. On the other hand, the tunability of System B allows the vehicle to be more under control (steerable) since the maximum value of deceleration requested can be set well below the friction limit. 

\subsubsection{Friction Retrieved from an Active Map}
In this last case, a more realistic scenario is presented. Considering the presence of an active map containing inaccurate ground friction estimates (due to several possible factors such as old data, measurements errors, ...). The value provided by the map ($\mu=0.6$) is higher than the real value ($\mu=0.4$).

\begin{figure}[h]
	\centering
\subfloat[]{
		\includegraphics[scale=0.275]{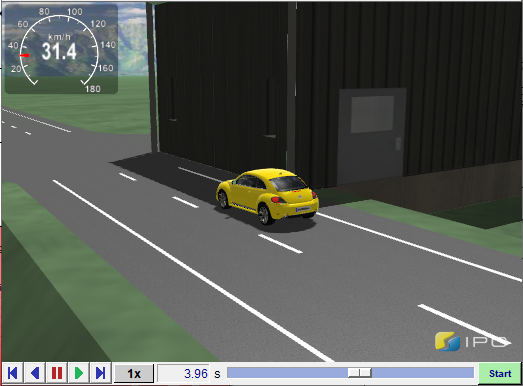}}
\subfloat[]{
		\includegraphics[scale=0.275]{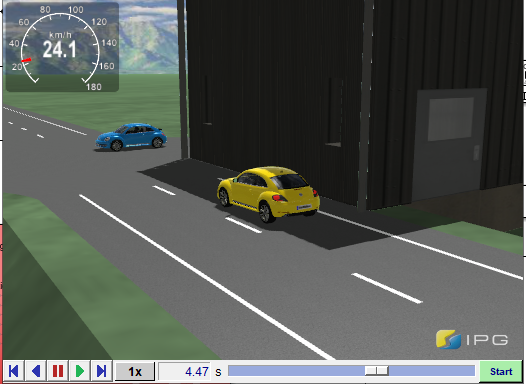}}
\subfloat[]{
		\includegraphics[scale=0.275]{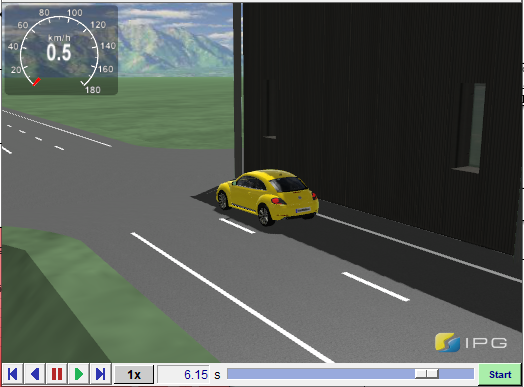}}

	\caption[Screenshots from CarMaker in Chronological Order]{Screenshots from CarMaker in Chronological Order: System Type B, Low Ground Friction}
	\label{fig:bdXd}
\end{figure}

\begin{figure}[h]
	\centering
	\subfloat[]{
		\includegraphics[scale=0.3]{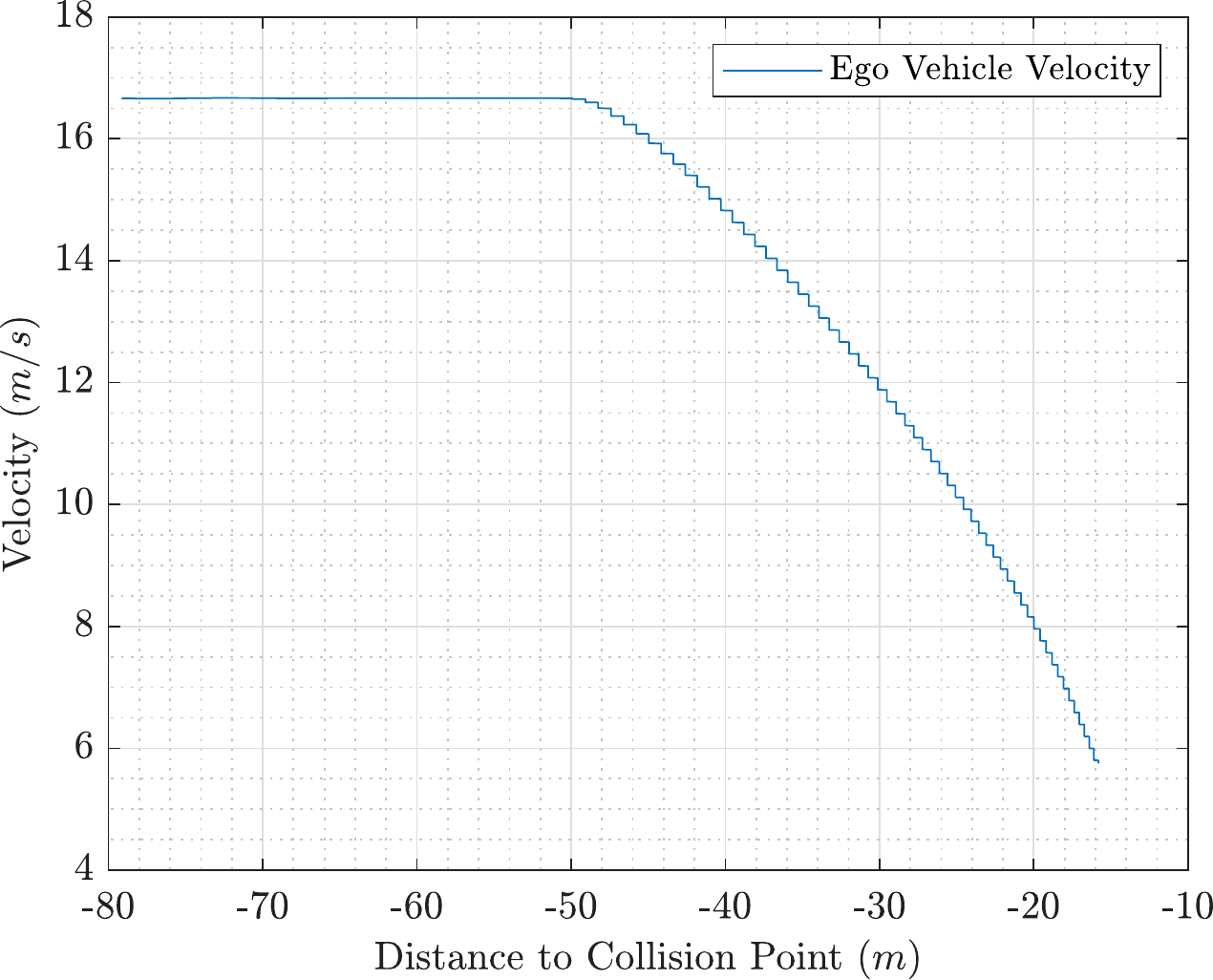}
		\label{fig:GDLF1}}
	\subfloat[]{	
		\includegraphics[scale=0.3]{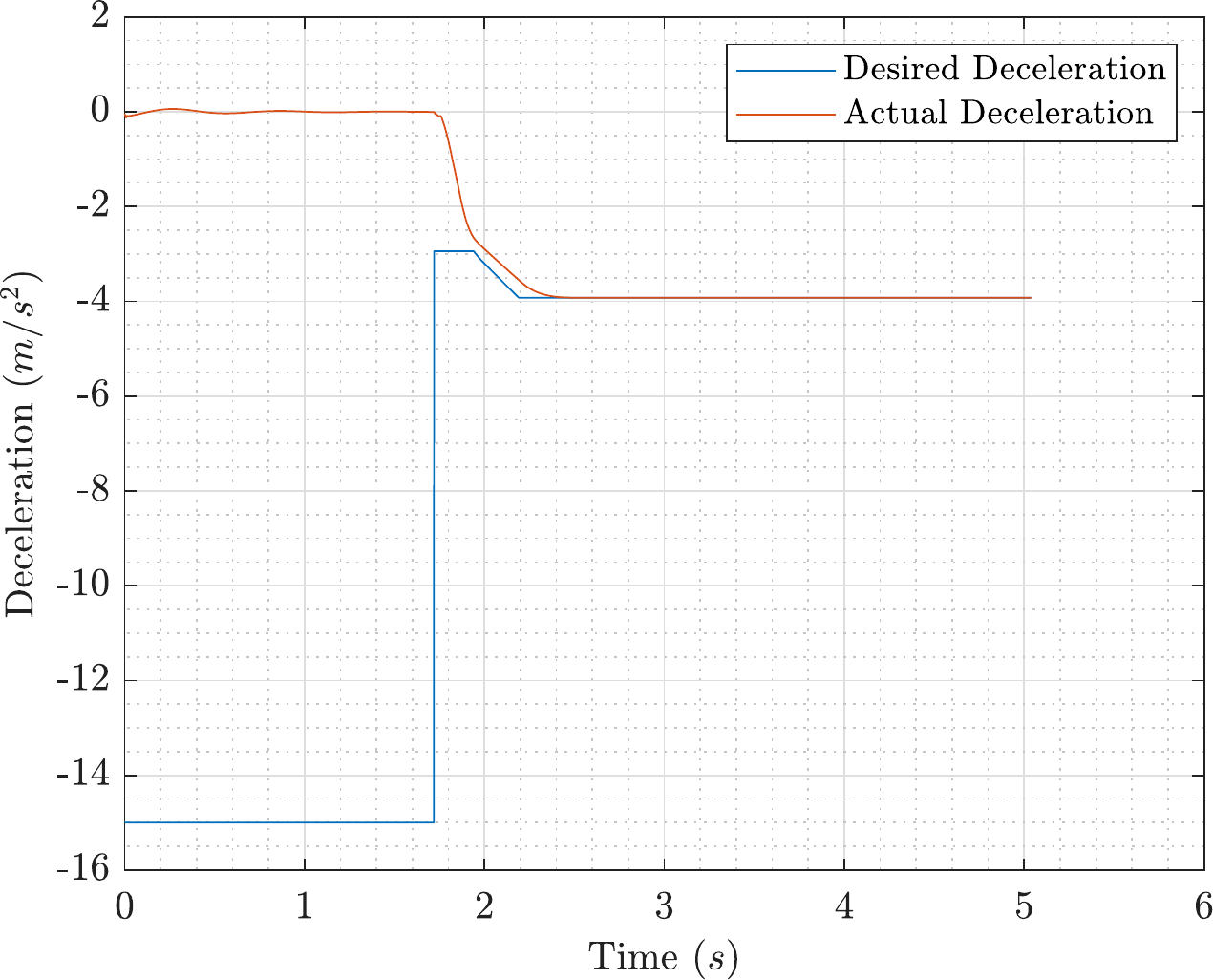}
		\label{fig:GDLF2}}\\
	\caption{System Type B: Low Ground Friction,(a) Desired Deceleration vs Actual Deceleration (b) Ego Vehicle Velocity vs Distance to Collision Point }
	\label{fig:newsx}
\end{figure}

Relevant numerical results are presented in Figure \ref{fig:newsx}. 

It could be seen that with respect to the latter situation (as in Figure \ref{fig:DLKF1}), the system starts braking later. However, the maximum possible deceleration is reached as fast as possible to ensure that the collision is avoided. It could be seen that the safety is improved as the system is now more adaptive to the environmental conditions.


\section{Conclusion}
To conclude, the aim of this work was the following:
\begin{itemize}
	\item develop and test two novel intersection collision avoidance systems that rely on information from 5G, radar sensors, and Pirelli CyberTyre; 	
	\item test the usability of 5G in safety systems of connected vehicles.
\end{itemize}

As presented in the paper, state of the art AEB systems usually issue a predefined braking request based on a static trigger. The advantage of such existing systems is the simplicity which reflects the simple hardware and reduced costs. Starting from these consideration, a novel approach consisting of an upgrade of the original architecture in order to limit costs of the design of new hardware and software, is presented. While keeping the same braking logic adopted in the radar system, an additional trigger was added to enable braking based on braking distance. The results of the virtual testing have been positive in both normal ground friction and low ground friction environments. In order to fully take advantage of the ground friction estimate, a second novel AEB control logic able to adapt itself during the braking is proposed. In detail, a braking scheme with an exponential-scheme was adopted as it ensures smooth braking as well as high robustness and adaptability to road conditions as shown by virtual simulations and comparison to the previous AEB proposed. However, due to the complex nature of the equations to be solved (to find $T$ and $\alpha$), the system requires much higher computational power. Also, the tuning of the parameter $a_\text{tuned}$ presents a disadvantage, as setting a high value will make the system start braking very early but the system will be more robust, and setting it very low will make system start braking \textit{at normal time} but this will limit robustness. 

\section*{Acknowledgments}
This work is done in collaboration between mechanical engineering department of Politecnico di Milano, iDrive Lab of Politecnico di Milano and Vodafone. The work presented is funded by, and part of the Vodafone 5G Trial in Milan.\newline
The authors would like to thank all their collaborating partners in the Vodafone 5G Trial in Milan: Altran, Vodafone Automotive, Magneti Marelli, Pirelli, and FCA. A special thanks goes to IPG for providing student licenses that were necessary for the development and testing of the developed work. A special thanks goes to Mr. Gianluca Stefanini for his help and advice on 5G communications.\newline
The Italian Ministry of Education, University and Research is acknowledged for the support provided through the Project ``Department of Excellence LIS4.0 - Lightweight and Smart Structures for Industry 4.0".

\section*{Disclosure statement}
	The author(s) declared no potential conflicts of interest with respect to the research, authorship, and/or publication of this article.

\section*{Funding}
	The author(s) disclosed receipt of the following financial support for the research, authorship, and/or publication of this article: This work has been supported by Vodafone Italia under Vodafone 5G Trial in Milan.

\end{document}